\documentclass[english,10pt, twocolumn, onespace]{IEEEtran}
\usepackage[T1]{fontenc}
\usepackage[latin9]{inputenc}
\usepackage{array}
\usepackage{float}
\usepackage{bm}
\usepackage{multirow}
\usepackage{amsthm}
\usepackage{amsmath}
\usepackage{amssymb}
\usepackage{graphicx}

\makeatletter

\providecommand{\tabularnewline}{\\}
\floatstyle{ruled}
\newfloat{algorithm}{tbp}{loa}
\providecommand{\algorithmname}{Algorithm}
\floatname{algorithm}{\protect\algorithmname}

 \let\oldforeign@language\foreign@language
 \DeclareRobustCommand{\foreign@language}[1]{%
   \lowercase{\oldforeign@language{#1}}}
\theoremstyle{plain}
\newtheorem{thm}{\protect\theoremname}
\theoremstyle{plain}
\newtheorem{lem}[thm]{\protect\lemmaname}
\theoremstyle{plain}
\newtheorem{cor}[thm]{\protect\corollaryname}

\usepackage{bm}
\usepackage{subfigure}
\usepackage{cite}
\usepackage[english]{babel}
\usepackage[T1]{fontenc}
\usepackage{algorithm}
\usepackage{algorithmic}
\vspace{-1in}

\makeatother

\usepackage{babel}
\providecommand{\corollaryname}{Corollary}
\providecommand{\lemmaname}{Lemma}
\providecommand{\theoremname}{Theorem}

\begin{document}

\title{Delay Optimal Scheduling \\ for Energy Harvesting Based Communications%
\thanks{J. Liu and H. Dai are with the Department of Electrical and Computer
Engineering, NC State University, Raleigh, NC 27695 (Email: hdai@ncsu.edu,
jliu23@ncsu.edu).  J. Liu and W. Chen are with State Key Laboratory
on Microwave and Digital Communications, Tsinghua National Laboratory
for Information Science and Technology (TNList) and Department of
Electronic Engineering, Tsinghua University, Beijing, China (e-mail:
eeliujuan@tsinghua.edu.cn, wchen@tsinghua.edu.cn).%
}}

\author{Juan Liu, Huaiyu Dai, \IEEEmembership{Senior Member, IEEE}, and
Wei Chen, \IEEEmembership{Senior Member, IEEE}}

\markboth{}{}
\maketitle
\begin{abstract}
Green communications attract increasing research interest recently.
Equipped with a rechargeable battery, a source node can harvest energy
from ambient environments and rely on this free and regenerative energy
supply to transmit packets. Due to the uncertainty of available energy
from harvesting, however, intolerably large latency and packet loss
could be induced, if the source always waits for harvested energy.
To overcome this problem, one Reliable Energy Source (RES) can be
resorted to for a prompt delivery of backlogged packets. Naturally,
there exists a tradeoff between the packet delivery delay and power
consumption from the RES. In this paper, we address the delay optimal
scheduling problem for a bursty communication link powered by a capacity-limited
battery storing harvested energy together with one RES. The proposed
scheduling scheme gives priority to the usage of harvested  energy,
and resorts to the RES when necessary based on the data and energy
queueing processes, with an average power constraint from the RES.
Through two-dimensional Markov chain modeling and linear programming
formulation, we derive the optimal threshold-based scheduling policy
together with the corresponding transmission parameters. Our study
includes three exemplary cases that capture some important relations
between the data packet arrival process and energy harvesting capability.
Our theoretical analysis is corroborated by simulation results. \end{abstract}

\section{Introduction}

\noindent Energy harvesting can provide renewable free energy supply
for wireless communication networks. With the help of solar cells,
thermoelectric and vibration absorption devices, and the like, communication
devices are able to gather energy from surrounding environments. 
Energy harvesting can also help reduce carbon emission and environmental
pollution, as well as the consumption of traditional energy resources
\cite{2003_AKansal_ISLPED,2003_MRahimi_HShah_ICRA,2009_WKGSeah_VITAE}.
In practice, harvested energy arrives in small units at random times
and the storage battery usually has limited capacity \cite{2010_JYang_CISS}.
Hence, wireless communication systems exclusively powered by energy
harvesting devices may not guarantee the users' quality of service.
To provide dependable communication service, reliable energy resources
can serve as backup in the case of energy shortage. In this way, efficient
mixed usage of the harvested energy and reliable energy provides a
key solution to robust wireless green communications \cite{2012_ZNiu_SZhou_TWC},
an emerging area of critical importance to future wireless development.

In wireless networks, energy efficient transmission has been an ever-present
important issue \cite{2002_AElGamal_TN,2005_MAZafer_infocom,2007_MNeely_2007}.
Subject to the randomness and causality of energy harvesting, the
optimal transmission problem has been investigated for an energy harvesting
wireless link with batteries of either finite or infinite capacity
in \cite{2010_JYang_CISS,2012_JYang_twc,2012_KTutuncuoglu_AYener_TWC}.
In these works, the authors assumed that the energy harvesting profile
($i.e.$, the arrival times and associated amount of harvested energy)
is known before the transmission starts. This line of work has been
extended to wireless fading channels \cite{2011_OOzel_JSAC}, broadcast
channels \cite{2011_JYang_ICC} and two-hop networks \cite{2012_OOrhan_EErkip}. 

Some other recent works have focused on developing efficient transmission
and resource allocation algorithms with different objectives and energy
recharging models. For example, a save-then-transmit protocol was
proposed in \cite{2012_RZhang_ISIT} to minimize the delay constrained
outage probability by using two alternating batteries, where the battery
charging rate is modeled as a random variable. In \cite{2010_MGatzianas_TWC},
a cross-layer resource allocation problem was studied for wireless
networks powered by rechargeable batteries, where the amount of replenished
energy is assumed to be independent and identically distributed in
each time slot. In \cite{2012_CKHo_RZhang_TSP}, an optimal energy
allocation problem was studied for a wireless link with time varying
channel conditions and energy sources. A line of work pertinent to
our study focuses on the queueing performance analysis for optimal
energy management policies. In particular, different sleep/wake-up
strategies in a solar-powered wireless sensor network were studied
in \cite{2007_DNiyato_EHossain_AFallahi_tmc}. Energy management policies
were proposed in \cite{2010_VSharma_twc} to maximize the stable throughput
and minimize the mean delay for energy harvesting sensor nodes. 

While a node can harvest an infinite amount of energy in the long
run, harvested energy actually arrives at random times. Due to the
energy causality constraint, the node should accumulate a sufficient
amount of energy before each packet transmission. Hence, the waiting
time could be undesirably long and some packets might be dropped due
to delay violation. Intuitively, this situation can be greatly relieved
if one Reliable Energy Source (RES) can be used to transmit backlogged
packets when needed. At the other extreme, the problem becomes trivial
if the system can always transmit using the reliable energy. Hence,
there exists a tradeoff between the packet delivery delay and the
energy consumption from the reliable source.

In this paper, we investigate the delay optimal scheduling policy
for a communication link powered by a capacity-limited battery storing
harvested energy and one RES.  In our system, the source will first
seek energy supply from the capacity limited energy harvesting battery
whenever available, and resort to the RES when necessary, but with
an average power constraint. In particular, subject to the bursty
energy harvesting profile, it transmits with one of the energy supplies
according to the data queue status and the energy storage status at
the battery. Under the constraint of the average power consumption
from the RES, we study the delay optimal scheduling problem, taking
into account the match and mismatch between the energy harvesting
capabilities and data packet arrival. 

To analyze the proposed scheme, we formulate a two-dimensional discrete-time
Markov chain and derive the steady-state probabilities. Based on the
Markov chain modeling, we can derive the average delay and the average
power consumed from the RES as functions of the steady-state probabilities.
Then, by formulating a Linear Programming (LP) problem and analyzing
its properties, we are able to characterize the structure of the optimal
solution. Moreover, we can obtain an elegant closed-form expression
for the optimal solution in the case where each unit of harvested
energy can support one data packet transmission. We also develop an
algorithm to find the optimal solutions in other cases. From the optimal
solution, we can determine the optimal probabilistic transmission
parameters. It is found that in the face of a depleted battery, the
optimal transmission strategy depends on a critical threshold for
the data queue length. In particular, the source relies on the harvested
energy supply if the data queue length is below the critical threshold,
and resorts to the RES otherwise. Our theoretical analysis is verified
 by simulations. 

The rest of this paper is organized as follows. Section \ref{sec:System-Model}
introduces the system model and the stochastic scheduling scheme.
In Section \ref{sec:Markov_chain}, a two-dimensional Markov chain
model is constructed for the data and energy packet queueing system.
Section \ref{sec:LP_problem} formulates an LP problem for our scheduling
objective. By analyzing the properties of the LP problem, we derive
the optimal steady-state probabilities and then determine the optimal
transmission parameters in Section \ref{sec:Delay-Optimal-Scheduling}.
Section \ref{sec:Simulation-Results} demonstrates the simulation
results and Section \ref{sec:Conclusions} concludes this paper. For
better illustration and in the interest of space, most proofs for
our results are put in the appendices.

\section{System Model\label{sec:System-Model}}

\begin{figure}[t]
\centering
\renewcommand{\figurename}{Fig.}\includegraphics[width=1\columnwidth]{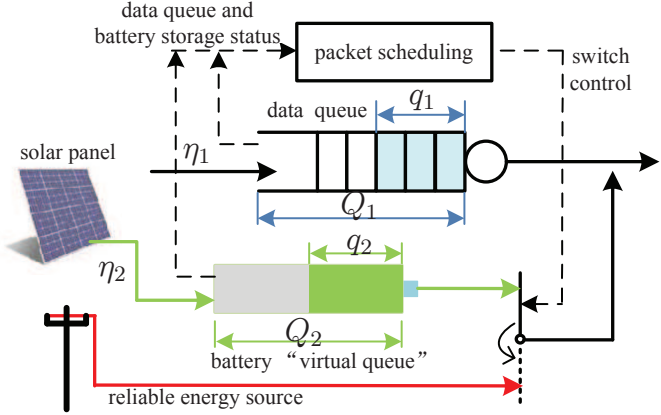}

\caption{System model. }
\label{fig:system_model}
\end{figure}

\subsection{System Description}

\noindent We consider a communication link which is powered mainly
by a battery storing the harvested energy and further by the RES when
necessary, as shown in Fig.$\,$\ref{fig:system_model}. The RES refers
to any reliable energy source, either traditional (such as power grid)
or newly developed. The source node (e.g. base station) employs a
buffer to store the backlogged packets randomly generated from higher-layer
applications. Suppose that the data packets arrive at the source buffer
according to a Bernoulli arrival process \cite{2000_TGRobertazzi_book}
with probability $\eta_{1}$. This simple yet widely adopted traffic
model allows tractable analysis, and provides insights for further
study. The system is assumed to be time-slotted, and at the beginning
instant of each slot, $k_{1}\in\mathbb{N}$ data packets arrive at
the data queue with capacity $Q_{1}$. In this work, $Q_{1}$ is treated
as sufficiently large (so no data overflow will incur) and fixed.
Let $q_{1}[t]\in\mathcal{Q}_{1}=\{0,1,2,\cdots,Q_{1}\}$ be the length
of the data queue at the end of slot $t$, updated as
\begin{equation}
q_{1}[t]=\min\{q_{1}[t-1]+a_{1}[t]-v_{1}[t],Q_{1}\},
\end{equation}
where $a_{1}[t]\in\{k_{1},0\}$ and $v_{1}[t]\in\{1,0\}$ denote the
number of data packets arriving and served in each time slot $t$,
respectively. Without loss of generality, it is assumed that at most
one packet is transmitted in each slot due to the capacity limitation
of the communication link. Extension to multi-packet transmission
will be considered in future work.

The harvested energy is generally sporadically and randomly available,
and we adopt a probabilistic energy harvesting model similar to \cite{2009_JLei_RYates_LGreenstein_twc}.
Assume that $e_{s}$ Joule harvested energy arrives at the beginning
of a time slot with probability $\eta_{2}$, which can be used to
transmit $k_{2}$ packets. That is $e_{s}=k_{2}\tilde{e}_{s}$, where
$\tilde{e}_{s}$ (Joule) denotes the amount of energy needed for transmission
of one data packet, and $k_{2}(\geq1)$ is rounded down to the nearest
integer. We will consider several interesting combinations of $k_{1}$
and $k_{2}$ in this study, and leave the case $k_{2}<1$ to future
study. The harvested energy is stored in the battery with the maximum
capacity $E$ Joule, and discarded when the battery is full. The battery
storage is modeled as an energy queue with a finite capacity $Q_{2}=\lfloor E/\tilde{e}_{s}\rfloor$,
where one unit of transmission energy $\tilde{e}_{s}$ is viewed as
one energy packet. Let $a_{2}[t]$ and $v_{2}[t]$ be the number of
energy packets received and consumed in each slot $t$, respectively.
At the end of time slot $t$, the length of the energy queue $q_{2}[t]\in\mathcal{Q}_{2}=\{0,1,2,\cdots,Q_{2}\}$
is updated as 
\begin{equation}
q_{2}[t]=\min\{q_{2}[t-1]+a_{2}[t]-v_{2}[t],Q_{2}\}.
\end{equation}
It is assumed that the packet and energy arrival processes are independent,
and the newly harvested energy can be used for data transmission in
the same slot. For notational convenience, we set $\bm{q}[t]=(q_{1}[t],q_{2}[t])$
to be the buffer status in the time slot $t$. Similarly, the arrival
and service processes can be characterized by the vectors $\bm{a}[t]=(a_{1}[t],a_{2}[t])$
and $\bm{v}[t]=(v_{1}[t],v_{2}[t])$, respectively.

\subsection{Stochastic Scheduling}

\noindent As we mentioned above, the source node is encouraged to
exploit the harvested energy whenever available, and resort to the
backup RES when necessary. To this end, the source should always transmit
using the energy stored in the battery or newly arriving energy packet
when possible, which corresponds to the case $q_{2}[t-1]>0$ or $a_{2}[t]>0$.
When the harvested energy is not available, $i.e.,\, q_{2}[t-1]=a_{2}[t]=0$,
the source schedules the transmission of data packets with the RES
energy according to the data queue status $q_{1}[t-1]$ and the data
packet arrival status $a_{1}[t]$. For generality, we define two sets
of parameters: $\{g_{i}\}$ and $\{f_{i}\}$ in our scheduling scheme.
In particular, with $q_{1}[t-1]=i$, if there is new data packet arrival
in this slot, $i.e.$, $a_{1}[t]>0$, the source node transmits one
data packet with probability $g_{i}$ with the RES energy and holds
from transmission with probability $1-g_{i}$, respectively; If no
new data packet arrives, $i.e.$, $a_{1}[t]=0$, it transmits with
probability $f_{i}$ and holds with probability $1-f_{i}$, respectively.
As discussed later, these parameters $\{g_{i}\}$ and $\{f_{i}\}$
shall be optimized to achieve the minimum average queueing delay in
different cases. 

According to the proposed scheduling policy, the service process $\bm{v}[t]$
depends on the queue status $\bm{q}[t-1]$ and the arrival process
$\bm{a}[t]$, as described below.
\begin{enumerate}
\item Case 1: $\bm{q}[t-1]=(0,j)$ ($j>0$) \\In this case, the source
can transmit a newly arriving data packet using the harvested energy
from the battery in the current time slot $t$, and the service process
can be expressed as
\begin{equation}
\bm{v}[t]=\begin{cases}
(1,1)\quad w.p.1, & \bm{a}[t]=(k_{1},\cdot),\\
(0,0) & \bm{a}[t]=(0,\cdot),
\end{cases}
\end{equation}
where $w.p.$ means 'with the probability'. The notation of $(k_{1},\cdot)$
is used to denote both $(k_{1},k_{2})$ and $(k_{1},0)$. 
\item Case 2: $\bm{q}[t-1]=(i,j)$ ($i>0$, $j>0$) \\In this case, the
source can transmit a backlogged packet with the harvested energy.
The service process is expressed as
\begin{equation}
\bm{v}[t]=(1,1)\quad w.p.1,\,\bm{a}[t]=(\cdot,\cdot).
\end{equation}

\item Case 3: $\bm{q}[t-1]=(0,0)$ \\ In this case, when both data and
energy packets arrive, the source will transmit with the energy harvested;
When new data packets arrive in the absence of energy harvesting,
the source shall use the energy from the RES to transmit with probability
$g_{0}$. Hence, the service process can be expressed as 
\begin{equation}
\bm{v}[t]=\begin{cases}
(1,1)\quad w.p.\,1 & \bm{a}[t]=(k_{1},k_{2})\\
(1,0)\quad w.p.\, g_{0} & \bm{a}[t]=(k_{1},0)\\
(0,0) & \text{otherwise}.
\end{cases}
\end{equation}

\item Case 4: $\bm{q}[t-1]=(i,0)$ ($i>0$) \\In this case, the source
will transmit definitely using the harvested energy if it is available
in the current slot $t$. Otherwise, it will transmit using the RES
energy with probability $g_{i}$ if $a_{1}[t]=k_{1}$ and with probability
$f_{i}$ if $a_{1}[t]=0$, respectively. The service process is characterized
as
\begin{equation}
\bm{v}[t]=\begin{cases}
(1,1)\quad w.p.\,1, & \bm{a}[t]=(\cdot,k_{2})\\
(1,0)\quad w.p.\, g_{i} & \bm{a}[t]=(k_{1},0)\\
(1,0)\quad w.p.\, f_{i} & \bm{a}[t]=(0,0).
\end{cases}
\end{equation}

\end{enumerate}
The above four cases include all possible scenarios.

\subsection{Average Delay and Power Consumption}

\noindent In a queueing system, the average queuing delay is an important
metric \cite{2002_RBerry_RGallager_TIT}. From the above description,
the queueing system can be modeled as a discrete-time Markov chain,
where each state represents the buffer status. Let $(i,j)$ be the
state that the data queue length is $i$ and the energy queue length
is $j$, and $\pi_{(i,j)}$ denote the steady-state probability of
state $(i,j)$. By the Little\textquoteright{}s law, the average queueing
delay is related to the average buffer occupancy, and can be computed
as 
\begin{equation}
\bar{D}=\frac{1}{k_{1}\eta_{1}}\sum\nolimits _{i=1}^{Q_{1}}i\pi_{i}=\frac{1}{k_{1}\eta_{1}}\sum\nolimits _{i=1}^{Q_{1}}i\sum\nolimits _{j=0}^{Q_{2}}\pi_{(i,j)},\label{eq:delay}
\end{equation}
where $\pi_{i}=\sum_{j=0}^{Q_{2}}\pi_{(i,j)}$ $(i,j\geq0)$.

\begin{figure*}[t]
\centering
\renewcommand{\figurename}{Fig.}\includegraphics[width=1\textwidth]{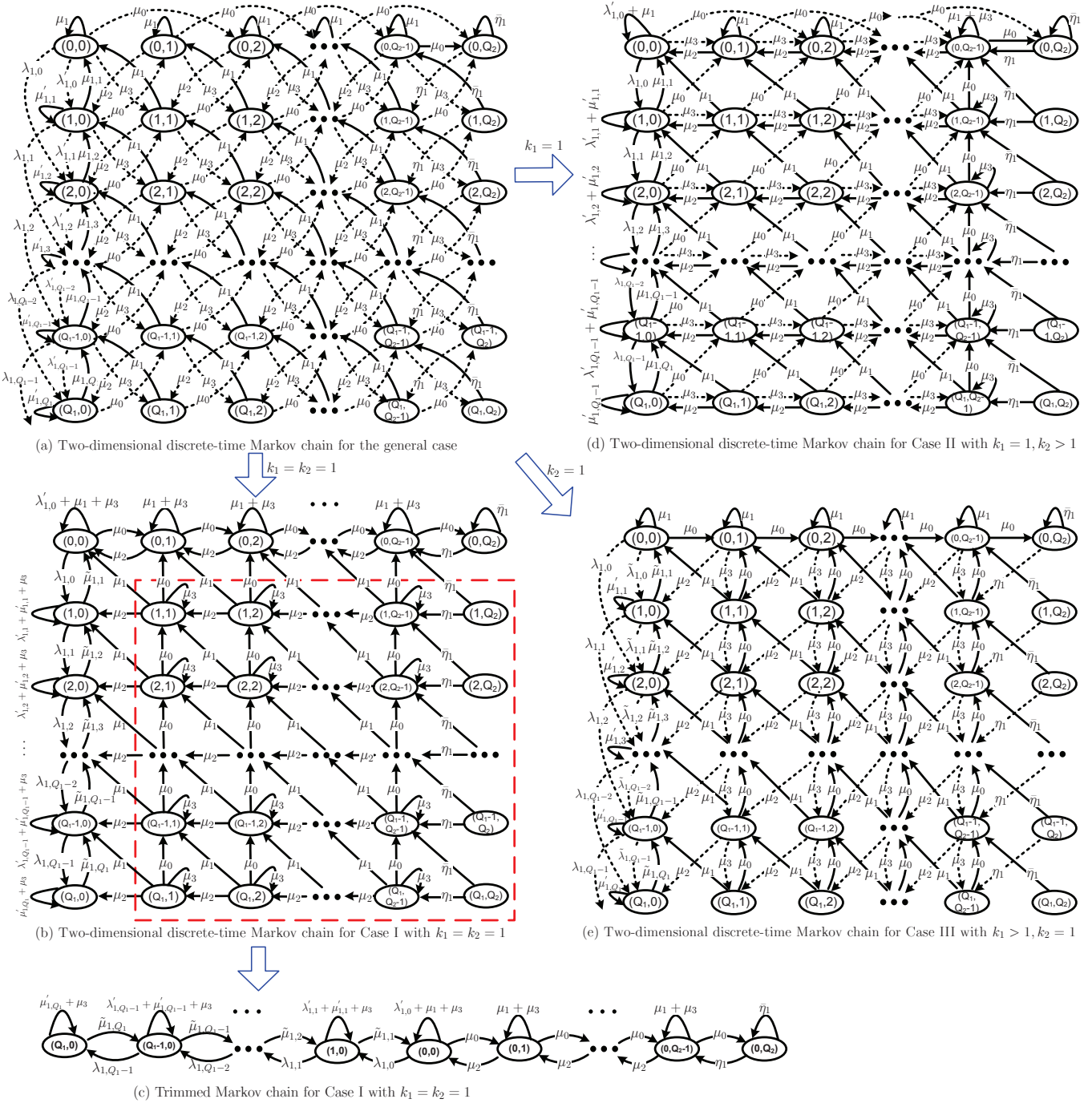}

\caption{Two-dimensional discrete-time Markov chain \protect\footnotemark. }
\label{fig:markov_chain_onefig}
\end{figure*}
The average transmission power is also an important performance metric
in wireless green communication systems. In this work, we focus on
the average power consumption from the RES. Denote by $c[t]$ the
power consumed in the $t$th time slot. If the source transmits using
the energy from the RES in time slot $t$, $c[t]=p:=\frac{1}{T_{s}}\tilde{e}_{s}$,
where $T_{s}$ denotes the transmission time. Otherwise, $c[t]=0$.
As will discussed below, the source draws one energy packet from the
RES depending on the current queueing status $\bm{q}[t]$. Let $\omega_{\bm{q}}(x)=\Pr\{c[t]=x|\bm{q}[t]\}$
denote the probability that the power consumption $c[t]$ is equal
to $x$ $(x\in\{0,p\})$ conditioned on the queue state $\bm{q}[t]$.
Using the law of total probability, we obtain the normalized average
power consumption (with respect to $p$) as
\begin{equation}
\bar{P}=\sum\nolimits _{\bm{q}\in\mathcal{Q}_{p}}\pi_{\bm{q}}\cdot\omega_{\bm{q}}(p),\label{eq:av_power}
\end{equation}
where $\mathcal{Q}_{p}$ is the set of states conditioned on which
the source may draw the RES energy to transmit one data packet. This
normalized quantity can be interpreted as the proportion of the number
of time slots in which the source transmits using the power from the
RES. From (\ref{eq:delay}) and (\ref{eq:av_power}), both the average
queueing delay and power consumption are functions of the steady-state
probabilities. In this work, we aim to study the delay optimal scheduling
policy which minimizes $\bar{D}$ subject to the average power constraint
$\bar{P}\leq p_{max}$ by determining the optimal transmission parameters
$\{g_{i}^{*}\}$ and $\{f_{i}^{*}\}$. As a key step, we will develop
two-dimensional Markov chain models for different combinations of
$k_{1}$ and $k_{2}$ in the next section.

\section{Two-dimensional Markov Chain Modeling\label{sec:Markov_chain}}

\noindent To analyze the proposed scheduling scheme, we formulate
a two-dimensional discrete-time Markov chain for the queueing system,
as shown in Fig.$\,$\ref{fig:markov_chain_onefig}.\footnotetext{The subfigure Fig.2(a) is intended for the general case of $k_1\geq1$ and $k_2\geq1$ (so the dashed lines are used for transitions); but $k_1=k_2=2$ can be assumed when checking the transition probabilities given in Section \ref{sec:Markov_chain}.}

Let $\Pr\{\bm{q}[t+1]|\bm{q}[t]\}$ denote the one-step transition
probability of the Markov chain, which is homogeneous by the scheme
description. For ease of expression, we define four constants as
\begin{equation}
\begin{split}\mu_{0}=(1-\eta_{1})\eta_{2}, & \quad\mu_{1}=(1-\eta_{1})(1-\eta_{2}),\\
\mu_{2}=\eta_{1}(1-\eta_{2}), & \quad\mu_{3}=\eta_{1}\eta_{2}.
\end{split}
\label{eq:constant_definition}
\end{equation}
We further define two subsets of $\mathcal{Q}_{i}$ as: $\mathcal{Q}_{i}^{L}=\{0,\cdots,Q_{i}-1\}$,
$\mathcal{Q}_{i}^{R}=\{1,\cdots,Q_{i}\}$, and set $\bar{\eta}_{i}=1-\eta_{i}$,
for $i=1,2$.

We now describe the one-step transition probabilities in Fig.$\,$\ref{fig:markov_chain_onefig}(a)
in detail, by grouping them into several types. We start with the
four transitions among each square unit, for example, those among
$(2,1)$, $(2,2)$, $(1,2)$ and $(1,1)$ in Fig.$\,$\ref{fig:markov_chain_onefig}(a).
First, let us examine the transition from $(2,1)$ to $(1,2)$, more
generally, from $(i,j)$ to $(i-1,\min\{j+k_{2},Q_{2}\}-1)$. This
corresponds to the case that there is no data but energy packet arrival,
and one backlogged data packet is delivered, so clearly the corresponding
probability is $\mu_{0}.$ When neither data nor energy packets arrive,
one data packet stored in the buffer can be transmitted using one
energy packet from the battery if there exists. In this case, the
state will transfer from $(i,j)$ to $(i-1,j-1)$ (e.g., from $(2,2)$
to $(1,1)$ in Fig.$\,$\ref{fig:markov_chain_onefig}(a)) with probability
$\mu_{1}$ for all $i>0$ and $Q_{2}>j>0$. When $k_{1}$ data packets
arrive while no energy is harvested, one data packet will be transmitted
using one energy packet if there is energy stored in the battery.
That is, the state will transfer from $(i,j)$ to $(i+k_{1}-1,j-1)$
(e.g., from $(1,2)$ to $(2,1)$ in Fig.$\,$\ref{fig:markov_chain_onefig}(a))
with probability $\mu_{2}$ for all $Q_{2}>j>0$. When data and energy
packets arrive simultaneously, one data packet is transmitted using
one energy packet. In this case, the state will transfer from $(i,j)$
to $(i+k_{1}-1,\min\{j+k_{2},Q_{2}\}-1)$ (e.g., from $(1,1)$ to
$(2,2)$ in Fig.$\,$\ref{fig:markov_chain_onefig}(a)) with probability
$\mu_{3}$ for $j\in Q_{2}^{L}$. 

The case $j=Q_{2}$ requires special treatment, as the battery is
full and the newly harvested energy has to be discarded anyway. With
the capacity limit in mind, we have $\Pr\{(i-1,Q_{2}-1)|(i,Q_{2})\}=\bar{\eta}_{1}$
for $i>0$, $\Pr\{(0,Q_{2})|(0,Q_{2})\}=\bar{\eta}_{1}$, and $\Pr\{(i+k_{1}-1,Q_{2}-1)|(i,Q_{2})\}=\eta_{1}$
for all $i$.

We then consider the first row in Fig.$\,$\ref{fig:markov_chain_onefig}(a).
When no data packets arrive and $k_{2}$ energy packets newly arrive,
the state $(0,j)$ will transfer to $(0,\min\{j+k_{2},Q_{2}\})$ with
the corresponding transition probability $\mu_{0}$ for $j\in\mathcal{Q}_{2}^{L}$.
We have mentioned that $\Pr\{(0,Q_{2})|(0,Q_{2})\}=\bar{\eta}_{1}$
is due to the capacity limitation of the battery. The state $(0,j)$
remains the same with probability $\mu_{1}$ (when neither data nor
energy packets arrive).

We now focus our attention on the group of transition probabilities
on the first column of Fig. 2, $\{\lambda_{1,i}\}$ and $\{\lambda_{1,i}^{'}\}$
($i\in\mathcal{Q}_{1}^{L}$), $\{\mu_{1,i}\}$ and $\{\mu_{1,i}^{'}\}$
($i\in\mathcal{Q}_{1}^{R}$), which corresponds to the case that there
is no storage of harvested energy in the battery, and can be obtained
as
\begin{equation}
\begin{split}\lambda_{1,i} & =\Pr\{(i+k_{1},0)|(i,0)\}=\mu_{2}(1-g_{i})\,(i\in\mathcal{Q}_{1}^{L}),\\
\lambda_{1,i}^{'} & =\Pr\{(i+k_{1}-1,0)|(i,0)\}=\mu_{2}g_{i}\,(i\in\mathcal{Q}_{1}^{L}),\\
\mu_{1,i} & =\Pr\{(i-1,0)|(i,0)\}=\mu_{1}f_{i}\,(i\in\mathcal{Q}_{1}^{R}),\\
\mu_{1,i}^{'} & =\Pr\{(i,0)|(i,0)\}=\mu_{1}(1-f_{i})\,(i\in\mathcal{Q}_{1}^{R}).
\end{split}
\label{eq:par_lambda1i_mu_1i_general}
\end{equation}
In particular, when $k_{1}$ data packets arrive while no energy is
harvested (which happens with probability $\mu_{2}$), $\lambda_{1,i}$
and $\lambda_{1,i}^{'}$ denote the transition probabilities from
state $(i,0)$ to $(i+k_{1},0)$ and $(i+k_{1}-1,0)$, respectively,
depending on whether one data packet is delivered with the reliable
energy in this slot (with probability $g_{i}$). When neither data
nor energy packets arrive (which happens with probability $\mu_{1}$),
$\mu_{1,i}$ and $\mu_{1,i}^{'}$ denote the transition probabilities
from state $(i,0)$ to $(i-1,0)$ and $(i,0)$, respectively, depending
on whether one data packet is transmitted using the reliable energy
 (with probability $f_{i}$).

We order the $N=(1+Q_{1})(1+Q_{2})$ states as $(0,0)$, $\cdots$,
$(0,Q_{2})$, $(1,0)$, $\cdots$, $(1,Q_{2})$, $\cdots$$(Q_{1},0)$,
$\cdots$, $(Q_{1},Q_{2})$, and let $\bm{P}$ denote the $N\times N$
transition matrix. We denote by $\bm{\pi}$ the $1\times N$ column
vector containing steady-state probabilities, and by $\bm{e}$ the
$N\times1$ column vector with all the elements equal to one. For
notational convenience, we also define two sub-vectors of $\bm{\pi}$
as: $\bm{\pi}_{i}=[\pi_{(i,0)};\cdots;\pi_{(i,Q_{2})}]$ and $\tilde{\bm{\pi}}_{i}=[\bm{\pi}_{0};\cdots;\bm{\pi}_{i}]$,
and denote by $\bm{e}_{i}$ a $1\times(i+1)(1+Q_{2})$ row vector
with all the elements equal to one. Given a set of parameters $\{g_{i}\}$
and $\{f_{i}\}$, the steady-state probabilities $\pi_{(i,j)}$ can
be obtained by solving the linear equations $\bm{\pi}\bm{P}=\bm{\pi}$
and $\bm{\pi}\bm{e}=1$. Note that the transmission parameters $\{g_{i}\}$
and $\{f_{i}\}$ only influence the transition probabilities from
the states $(i,0)$, $i\in Q_{1}$. We thus consider $\tilde{\bm{P}}_{s}$,
a submatrix of $\tilde{\bm{P}}=\bm{P}-\bm{I}$, to exclude the state
transition starting from states $(i,0)$. In this way, $\bm{\pi}\tilde{\bm{P}}_{s}=\bm{0}$
present the local balance equations at the states $(i,j)$ $(i\geq0,j>0)$.
For ease of expression, we also denote by $\tilde{\bm{P}}_{i}$ the
left-top submatrix of $(i+1)Q_{2}$ dimensions from $\tilde{\bm{P}}_{s}$.

In the general case with $k_{1}\geq1$ and $k_{2}\geq1$, the corresponding
Markov chain seems not amenable to analysis. In this paper, we mainly
focus on three cases: Case I with $k_{1}=1$ and $k_{2}=1$, Case
II with $k_{1}=1$ and $k_{2}>1$, and Case III with $k_{1}>1$ and
$k_{2}=1$, respectively. These three exemplary cases nonetheless
capture some important relations between the data and energy arrival
processes, and serve as the basis for further extensions. In the following,
we illustrate the Markov chain for each of the three cases.

\subsection{Case I: $k_{1}=1$ and $k_{2}=1$}

\noindent In this case, one data packet and one energy packet arrive
in each slot with probabilities $\eta_{1}$ and $\eta_{2}$, respectively.
Accordingly, the simplified Markov chain is shown in Fig.$\,$\ref{fig:markov_chain_onefig}(b).
Essentially all expressions in the general case carry over with the
substitution of $k_{1}=k_{2}=1$. For example, the transition from
$(i,j)$ to $(i-1,\min\{j+k_{2},Q_{2}\}-1)$ in Fig.$\,$\ref{fig:markov_chain_onefig}(a)
becomes that from $(i,j)$ to $(i-1,j)$ in Fig.$\,$\ref{fig:markov_chain_onefig}(b),
again with probability $\mu_{0}$. This applies to the states in the
first column as well, and as a result, a new notation is needed for
the transition from $(i,0)$ to $(i-1,0)$, which combines $\mu_{0}$
and the previous $\mu_{1,i}$:
\begin{equation}
\begin{split}\tilde{\mu}_{1,i} & =\Pr\{(i-1,0)|(i,0)\}=\mu_{0}+\mu_{1}f_{i}\end{split}
\label{eq:par_mu_1i_case1}
\end{equation}
for all $i\in\mathcal{Q}_{1}^{R}$. Also, it is worth noting that
in the dashed square, neither queue length can ever increase regardless
of the arrival processes, as one data packet transmission happens
for sure. As a result, the states $\bm{q}[t]$ with $q_{1}[t]\cdot q_{2}[t]>0$
are transient in the following lemma.  
\begin{lem}
\label{lem:recurrent_transient_states}In Case I with $k_{1}=k_{2}=1$
when $\eta_{1}<1$ or $\eta_{2}<1$, the queue status satisfying $q_{1}[t]\cdot q_{2}[t]>0$
is transient. \end{lem}
\begin{IEEEproof}
Let $f_{\bm{q}}^{(n)}$ denote the probability that the queue state
$\bm{q}[t]$ will return to itself for the first time after $n$ steps.
As shown in Fig.$\,$\ref{fig:markov_chain_onefig}(b), when $i\cdot j>0$,
$f_{\bm{q}}^{(1)}=\Pr\{(i,j)|(i,j)\}=\mu_{3}$ and $f_{\bm{q}}^{(n)}=0$
for $n>1$. Hence, $\sum_{n=1}^{\infty}f_{\bm{q}}^{(n)}=\mu_{3}=\eta_{1}\eta_{2}<1$,
when $\eta_{1}<1$ or $\eta_{2}<1$. From \cite{1996_ANShiryaev_probability},
the state $\bm{q}[t]$ with $q_{1}[t]\cdot q_{2}[t]>0$ is a transient
state. 
\end{IEEEproof}
\emph{} This implies that either the data queue or the energy queue
will be exhausted, even if they are not empty initially. Hence, when
calculating the steady-state probabilities $\pi_{(i,j)}$, the two-dimensional
Markov chain can be reduced to the one-dimensional one, as plotted
in Fig.$\,$\ref{fig:markov_chain_onefig}(c), which consists of the
states $(i,0)$ and $(0,j)$ for all $i\in\mathcal{Q}_{1}$ and $j\in\mathcal{Q}_{2}$.

\subsection{Case II and Case III }

\noindent In Case II, $k_{2}$ energy packets ($k_{2}>1$) arrive
at the battery with probability $\eta_{2}$ per slot. Hence, the length
of the energy queue may increase by $k_{2}$ or $k_{2}-1$ (when one
energy packet is consumed in the current slot) each time. The resulting
two-dimensional Markov chain is shown in Fig.$\,$\ref{fig:markov_chain_onefig}(d).
In Case III, $k_{1}>1$ data packets arrive with the probability $\eta_{1}$
at each slot, and the two-dimensional Markov chain is illustrated
in Fig.$\,$\ref{fig:markov_chain_onefig}(d), where the data queue
length could increase by $k_{1}$ or $k_{1}-1$ (when one data packet
is transmitted using an energy packet harvested or drawn from the
RES in the current slot). 

As shown in Fig.$\,$\ref{fig:markov_chain_onefig}(d), the solid
lines present the fixed state transitions while the dotted lines indicate
state transitions that vary with different $k_{2}$. In particular,
the state $(i,j)$ transfers to $(i-1,\min\{j+k_{2},Q_{2}\}-1)$ with
the probability $\mu_{0}$ and to $(i,\min\{j+k_{2},Q_{2}\}-1)$ with
the probability $\mu_{3}$, respectively. Similarly, the state $(0,j)$
transfers to $(0,\min\{j+k_{2},Q_{2}\})$ with the probability $\mu_{0}$,
and to $(0,\min\{j+k_{2},Q_{2}\}-1)$ with the probability $\mu_{3}$,
respectively. Note that the states $(i,Q_{2})$ for all $i>0$ are
transient. 

Similarly in Fig.$\,$\ref{fig:markov_chain_onefig}(e), solid and
dotted lines are used to present the fixed state transitions and state
transitions that vary with different $k_{1}$, respectively. Similar
to Case I, the state transfers from $(i,0)$ to $(i-1,0)$ with the
combined transition probability $\tilde{\mu}_{1,i}=\mu_{0}+\mu_{1}f_{i}$.
For the same reason, the transition probability from $(i,0)$ to $(i+k_{1}-1,0)$
is
\begin{equation}
\tilde{\lambda}_{1,i}=\mu_{3}+\mu_{2}g_{i}=\mu_{3}+\lambda_{1,i}^{'}=\eta_{1}-\lambda_{1,i}.\label{eq:par_lambda_case3}
\end{equation}
And the states $(i,Q_{2})$ for all $i>0$ are transient.

\section{LP Problem Formulation\label{sec:LP_problem}}

\noindent As discussed above, both the average delay and power consumption
from the RES are functions of the steady-state probabilities of the
corresponding Markov chains, which in turn depend on the transmission
parameters $\{g_{i}\}$ and $\{f_{i}\}$ to be designed. To seek the
optimal scheduling policy, we adopt a two-step procedure \cite{2012_WChen_TNET}:
first we formulate an LP problem only depending on the steady-state
probabilities, and obtain the corresponding solution; then from the
optimal solution of the LP problem, we determine the optimal transmission
parameters.

Our objective is to minimize the average queueing delay subject to
the maximum average power constraint from the RES. The corresponding
LP problem can be formulated as 
\begin{equation}
\begin{split}\min & \quad\bar{D}=\frac{1}{k_{1}\eta_{1}}\sum{}_{i=1}^{Q_{1}}i\sum{}_{j=0}^{Q_{2}}\pi_{(i,j)}\\
s.t. & \begin{cases}
\bar{P}=\sum\limits _{i=0}^{Q_{1}}\xi_{i}\cdot\pi_{(i,0)}-\sum\limits _{i=0}^{Q_{1}}\zeta_{i}\cdot\pi_{(i,1)}\leq p_{max}, & (a)\\
\Theta_{l}(i,\tilde{\bm{\pi}}_{i-1})\leq\sum\limits _{j=0}^{Q_{2}}\pi_{(i,j)}\leq\Theta_{u}(i,\tilde{\bm{\pi}}_{i})(i>0), & (b)\\
\pi_{(i,j)}\geq0,\,(\forall i,j), & (c)\\
\sum{}_{i=0}^{Q_{1}}\sum{}_{j=0}^{Q_{2}}\pi_{(i,j)}=1, & (d)\\
\bm{\pi}\tilde{\bm{P}}_{s}=\bm{0}. & (e)
\end{cases}
\end{split}
\label{eq:LP_problem}
\end{equation}
From the properties of a Markov chain, the last three constraints
(c)-(e) are straightforward. The original definition of $\bar{P}$
(c.f. (\ref{eq:av_power})) in constraint (a) does depend on the transmission
parameters; to facilitate derivation, we will give a new expression
for $\bar{P}$ in Lemma \ref{lem:ave_power} below that is only a
function of the steady-state probabilities $\pi_{(i,0)}$ and $\pi_{(i,1)}$,
$i\in Q_{1}$. The influence of the transmission parameters on the
problem is encapsulated in  the constraint (b), which represents the
relationship between the steady-state probabilities $\pi_{(i,j)}$
due to the varying transmission parameters $\{g_{i}\}$ and $\{f_{i}\}$,
as discussed later in Lemma \ref{lem:relation_pi}.  The optimal solution
to (\ref{eq:LP_problem}) is denoted by $\pi_{(i,j)}^{*}$ and the
minimum average delay by $\bar{D}^{*}$. 
\begin{table*}[t]
\centering
\renewcommand{\arraystretch}{1.5} 

\caption{The coefficients $\xi_{i}$ and $\zeta_{i}$ for Cases I, II, and
III.\label{Table_coefficients}}

\begin{tabular}{|c|l|l|l|}
\hline 
 & Case I with $k_{1}=k_{2}=1$ & Case II with $k_{1}=1$ and $k_{2}>1$ & Case III with $k_{1}>1$ and $k_{2}=1$\tabularnewline
\hline 
\multirow{2}{*}{$\xi_{i}$} & $\xi_{0}=\mu_{2}$ & \multirow{3}{*}{$\xi_{i}=\mu_{2}+\eta_{2}(Q_{1}-i)\,(i\in\mathcal{Q}_{1})$} & $\xi_{0}=\mu_{0}Q_{1}-\mu_{3}+k_{1}\eta_{1}$\tabularnewline
\cline{2-2} \cline{4-4} 
 & $\xi_{i}=\mu_{2}-\mu_{0}(i\in\mathcal{Q}_{1}^{R})$ &  & $\xi_{i}=k_{1}\eta_{1}-\eta_{2}\,(1\leq i\leq Q_{1}-k_{1})$\tabularnewline
\cline{4-4} 
 &  &  & $\xi_{i}=\eta_{1}(Q_{1}-i)-\eta_{2}\,(Q_{1}-k_{1}+1\leq i\leq Q_{1})$\tabularnewline
\hline 
\multirow{2}{*}{$\zeta_{i}$} & \multirow{3}{*}{$\zeta_{i}=0\,(i\in\mathcal{Q}_{1})$} & $\zeta_{0}=\mu_{2}Q_{1}$ & $\zeta_{0}=\mu_{2}(Q_{1}-k_{1}+1)$\tabularnewline
\cline{3-4} 
 &  & \multirow{2}{*}{$\zeta_{i}=\bar{\eta}_{2}(Q_{1}-i)+\mu_{1}\,(i\in\mathcal{Q}_{1}^{R})$} & $\zeta_{i}=(\mu_{1}+\mu_{2})(Q_{1}+1-i)-\mu_{2}k_{1}\,(1\leq i\leq Q_{1}-k_{1})$\tabularnewline
\cline{4-4} 
 &  &  & $\zeta_{i}=\mu_{1}(Q_{1}+1-i)\,(Q_{1}-k_{1}+1\leq i\leq Q_{1})$\tabularnewline
\hline 
\end{tabular}
\end{table*}
\begin{table*}
\centering
\renewcommand{\arraystretch}{1.5} 

\caption{$\Theta_{u}(i,\tilde{\bm{\pi}}_{i})$ and $\Theta_{l}(i,\tilde{\bm{\pi}}_{i-1})$
for Cases I, II, and III.\label{Table_upper_lower_bounds}}

\begin{tabular}{|c|c|c|c|c|}
\hline 
 & Case I with $k_{1}=k_{2}=1$ & Case II with $k_{1}=1,k_{2}>1$ & \multicolumn{2}{c|}{Case III with $k_{1}>1,k_{2}=1$}\tabularnewline
\hline 
\multirow{2}{*}{$\Theta_{u}(i,\tilde{\bm{\pi}}_{i})$} & \multirow{2}{*}{$\phi\pi_{(i-1,0)}$} & \multirow{2}{*}{$\tau\bar{\eta}_{2}\pi_{(i-1,0)}+\pi_{(i,0)}\bar{\eta}_{2}$} & $i<k_{1}$ & $\pi_{(i,0)}\bar{\eta}_{2}+\sum\limits _{m=0}^{i-1}\sum\limits _{j=0}^{Q_{2}}\tau\pi_{(m,j)}$\tabularnewline
\cline{4-5} 
 &  &  & $i\geq k_{1}$ & $\tau\bar{\eta}_{2}\pi_{(i-k_{1},0)}+\pi_{(i,0)}\bar{\eta}_{2}+\sum\limits _{m=i-k_{1}+1}^{i-1}\sum\limits _{j=0}^{Q_{2}}\tau\pi_{(m,j)}$\tabularnewline
\hline 
$\Theta_{l}(i,\tilde{\bm{\pi}}_{i-1})$ & $0$ & $0$ & \multicolumn{2}{c|}{$\sum\limits _{m=[i-k_{1}+1]^{+}}^{i-1}\sum\limits _{j=0}^{Q_{2}}\tau\pi_{(m,j)}$}\tabularnewline
\hline 
\end{tabular}
\end{table*}

\begin{lem}
\label{lem:ave_power}In Cases I, II and III, the normalized average
power consumption from the RES can be expressed as
\begin{equation}
\bar{P}=\sum{}_{i=0}^{Q_{1}}\xi_{i}\cdot\pi_{(i,0)}-\sum{}_{i=0}^{Q_{1}}\zeta_{i}\cdot\pi_{(i,1)},\label{eq:ave_power_cof}
\end{equation}
where the coefficients $\xi_{i}$ and $\zeta_{i}$ are presented in
Table \ref{Table_coefficients}.\end{lem}
\begin{IEEEproof}
The proof is deferred to Appendix \ref{sub:Proof-of-Lemma2}.
\end{IEEEproof}
\emph{Remark}: By exploiting the local balance equations of states
$(i,0)$ $(i\in Q_{1}^{L})$, we can replace all the items $\pi_{(i,0)}\mu_{2}g_{i}$
$(i\in\mathcal{Q}_{1})$ and $\pi_{(i,0)}\mu_{1}f_{i}$ $(i\in\mathcal{Q}_{1}^{R})$
of $\bar{P}$ with the items $\xi_{i}\pi_{(i,0)}$ and $\zeta_{i}\pi_{(i,1)}$
$(i\in\mathcal{Q}_{1})$. In this way, the average power consumption
$\bar{P}$ becomes a linear function of the steady-state probabilities
$\pi_{(i,0)}$ and $\pi_{(i,1)}$. Thus, the direct dependence of
$\bar{P}$ on the transmission parameters $\{g_{i}\}$ and $\{f_{i}\}$
is removed. 

Then, we discuss the constraint (\ref{eq:LP_problem}.b). The basic
idea is to vary the transmission parameters $\{g_{i}\}$ and $\{f_{i}\}$
in the full range of $[0,1]$, so as to obtain an upper and lower
bound for each $\pi_{i}$. In this way, we transform the constraints
on $\{g_{i}\}$ and $\{f_{i}\}$ into the relationship between the
steady-state probabilities themselves, which allows us to obtain the
optimal solution to (\ref{eq:LP_problem}) in terms of $\{\pi_{(i,j)}\}$
first. For ease of illustration, we define several constants as $\tau=\frac{\eta_{1}}{1-\eta_{1}}$,
$\phi=\frac{\mu_{2}}{\mu_{0}}$, and $\phi_{1}=\frac{\eta_{1}}{\mu_{0}}$.
Let us define $[x]^{+}=\max\{0,x\}$.
\begin{lem}
\label{lem:relation_pi}In Cases I, II and III, the probability $\pi_{i}$
satisfies
\begin{equation}
\Theta_{l}(i,\tilde{\bm{\pi}}_{i-1})\leq\pi_{i}=\sum{}_{j=0}^{Q_{2}}\pi_{(i,j)}\leq\Theta_{u}(i,\tilde{\bm{\pi}}_{i})\,(i>0),\label{eq:bounds_pi}
\end{equation}
where $\Theta_{u}(\cdot)$ and $\Theta_{l}(\cdot)$ are presented
in Table \ref{Table_upper_lower_bounds}. \end{lem}
\begin{IEEEproof}
The proof is deferred to Appendix \ref{sub:Proof-of-Lemma3}.
\end{IEEEproof}
\emph{Remark}: From the proof of Lemma \ref{lem:relation_pi}, we
have $\pi_{i}=\Theta_{u}(i,\tilde{\bm{\pi}}_{i})$ at $g_{i-k_{1}}=f_{i}=0$,
and $\pi_{i}=\Theta_{l}(i,\tilde{\bm{\pi}}_{i-1})$ at $g_{i-k_{1}}=f_{i}=1$\textsl{,
}\textsl{\emph{respectively, in all the three cases}}\textsl{ }%
\footnote{\textsl{\emph{More rigorously, in Case I, $\pi_{i}=\Theta_{l}(i,\tilde{\bm{\pi}}_{i-1})$
holds just when $g_{i-1}=1$ and $f_{i}$ can be arbitrary.}}\textsl{ }%
}\textsl{\emph{. This lies in the fact that the transmission parameters
$\{g_{i}\}$ and $\{f_{i}\}$ determine the relationship between }}the
steady-state probabilities\textsl{\emph{ $\{\pi_{(i,j)}\}$, and vice
versa. }}As listed in Table \ref{Table_upper_lower_bounds}, $\Theta_{u}(i,\tilde{\bm{\pi}}_{i})$
is a linear function of the steady-state probabilities $\pi_{(i-k_{1},0)},\cdots,\pi_{(i-1,Q_{2})},\pi_{(i,0)}$,
and $\Theta_{l}(i,\tilde{\bm{\pi}}_{i-1})$ is a linear function of
$\pi_{([i-k_{1}+1]^{+},0)},\cdots,\pi_{(i-1,Q_{2})}$.  

From Lemmas \ref{lem:ave_power} and \ref{lem:relation_pi}, $\bar{P}$,
$\Theta_{u}(i,\tilde{\bm{\pi}}_{i})$ and $\Theta_{l}(i,\tilde{\bm{\pi}}_{i-1})$
are all linear functions of the steady-state probabilities $\{\pi_{(i,j)}\}$.
Hence, we can represent them in the form of $\bar{P}(\bm{\pi})=\bm{\pi}\bm{a}_{0}$,
$\sum\nolimits _{j=0}^{Q_{2}}\pi_{(i,j)}-\Theta_{u}(i,\tilde{\bm{\pi}}_{i})=\bm{\pi}\bm{a}_{i}^{u}\,(i>0)$,
and $\Theta_{l}(i,\tilde{\bm{\pi}}_{i-1})-\sum\nolimits _{j=0}^{Q_{2}}\pi_{(i,j)}=\bm{\pi}\bm{a}_{i}^{l}\,(i>0)$,
where $\bm{a}_{0}$, $\bm{a}_{i}^{u}$ and $\bm{a}_{i}^{l}$ are $N\times1$
column vectors collecting corresponding coefficients.

\section{Delay Optimal Scheduling Under Power Constraint\label{sec:Delay-Optimal-Scheduling}}

\noindent In this section, we discuss the optimal solution to Problem
(\ref{eq:LP_problem}) by studying its structure with respect to the
steady-state probabilities of the corresponding Markov chains.

\subsection{Structure of The Optimal Solution}

\noindent For ease of discussion, we first consider a scheduling
policy strictly based on the threshold $m$: the source waits for
the harvested energy when the number of backlogged data packets is
\emph{less than or equal to} a certain threshold $m$ and transmits
using the reliable energy when the data queue length exceeds $m$.
According to the threshold $m$, we use $\tilde{p}_{m}$ to measure
the amount of power drawn from the RES. Since $\tilde{p}_{m}$ is
sufficient for the application of the scheduling policy based on the
threshold $m+1$, but not vise versa, $\tilde{p}_{m}$ is non-increasing
with the threshold $m$. We will show that the threshold based scheduling
policy turns out to be the optimal and the optimal threshold is determined
by the power thresholds $\{\tilde{p}_{m}\}$. 
\begin{thm}
\label{thm:opt_threshold} The optimal threshold is $i^{*}=0$ when
$p_{max}\geq\tilde{p}_{0}$, and $i^{*}>0$ when $k_{1}\eta_{1}-k_{2}\eta_{2}<p_{max}<\tilde{p}_{0}$,
respectively. \end{thm}
\begin{IEEEproof}
The proof is deferred to Appendix \ref{sub:Proof-of-Theorem4}.
\end{IEEEproof}
 We notice that the average queueing delay $\bar{D}=\frac{1}{k_{1}\eta_{1}}\sum_{i=1}^{Q_{1}}i\pi_{i}$
is a weighted summation of the steady-state probabilities $\pi_{i}$.
Thus, $\bar{D}$ can be reduced, if we assign a larger value to $\pi_{i}$
with a smaller index $i$ and vice versa. Based on this intuition,
we can reveal that the optimal solution to the LP problem (\ref{eq:LP_problem})
corresponds to a threshold based scheduling policy with the optimal
threshold $i^{*}$ determined by the maximum allowable power consumption
from the RES $p_{max}$. 
\begin{thm}
\label{thm:opt_structure}The optimal solution $\bm{\pi}^{*}$ satisfies
\begin{equation}
\begin{split}\bm{\pi}^{*}\bm{a}_{0} & \leq p_{max},\\
\bm{\pi}^{*}\bm{a}_{i}^{u} & =0\,(i=1,\cdots,i^{*}-1),\\
\bm{\pi}^{*}\bm{a}_{i}^{l} & =0\,(i=i^{*}+1,\cdots,Q_{1}),
\end{split}
\label{eq:length_condition_for_optimality}
\end{equation}
where the optimal threshold is obtained as
\begin{equation}
i^{*}=\arg\min_{\tilde{p}_{m}\leq p_{max}}m.
\end{equation}
 \end{thm}
\begin{IEEEproof}
The proof is deferred to Appendix \ref{sub:Proof-of-Theorem6}.
\end{IEEEproof}
\emph{Remark}: According to Lemma \ref{lem:relation_pi}, we have
$\pi_{i}=\Theta_{u}(i,\tilde{\bm{\pi}}_{i})$ or $\bm{\pi}\bm{a}_{i}^{u}=0$
when $g_{i-k_{1}}=f_{i}=0$, and $\pi_{i}=\Theta_{l}(i,\tilde{\bm{\pi}}_{i-1})$
when $g_{i-k_{1}}=f_{i}=1$, respectively. Therefore, associated with
(\ref{eq:length_condition_for_optimality}) is a threshold based scheduling
policy that waits for the harvested energy when the number of backlogged
data packets is \emph{less than} a certain threshold $i^{*}$, and
draws the reliable energy definitely when the harvested energy is
not available while the number of backlogged data packets exceeds
the threshold ($i^{*}$ if there is no new data packet arrival, and
$i^{*}-k_{1}$ if there is new data packet arrival).

Note that the LP problem (\ref{eq:LP_problem}) has an optimal solution
only when the queueing system is stable, $i.e.$, when the service
rate is greater than the arrival rate, according to Loynes\textquoteright{}s
theorem \cite{1962_RMLoynes_stability}. Throughout this paper, the
service rate is specialized as the total amount of energy that can
be drawn either from the RES or from the battery, $p_{max}+k_{2}\eta_{2}$.
 Hence, we will discuss the optimal solution to the LP problem (\ref{eq:LP_problem})
under the assumption that $p_{max}>k_{1}\eta_{1}-k_{2}\eta_{2}$.

\subsection{The Optimal Solution}

\noindent By exploiting the result in Theorem \ref{thm:opt_structure},
we continue to derive the optimal steady-state probabilities for Case
I, and develop an algorithm to obtain the optimal solutions for Case
II and Case III, respectively.

\subsubsection{Case I}

In this case, the two-dimensional Markov chain is reduced to a one-dimensional
one, where transitions takes place only between adjacent states, as
shown in Fig.$\,$\ref{fig:markov_chain_onefig}(c). We only need
to discuss the optimal steady-state probabilities $\pi_{(i,0)}^{*}$
and $\pi_{(0,j)}^{*}$ for all $i\in\mathcal{Q}_{1}$ and $j\in\mathcal{Q}_{2}$.
In the sequel, we first show that the optimal $\pi_{(0,j)}^{*}$ is
a function of $\pi_{(0,0)}^{*}$ in Lemma \ref{lem:pi_2j} and then
present the optimal $\pi_{(i,0)}^{*}$ in Corollary \ref{cor:opt_solution_SA}. 
\begin{lem}
\label{lem:pi_2j}In Case I, the optimal steady-state probability
$\pi_{(0,j)}^{*}$ is related to $\pi_{(0,0)}^{*}$ as 
\begin{equation}
\begin{split}\pi_{(0,j)}^{*} & =\begin{cases}
\pi_{(0,0)}^{*}\phi^{-j}, & 1\leq j\leq Q_{2}-1,\\
\pi_{(0,0)}^{*}\phi^{-(Q_{2}-1)}\phi_{1}^{-1}, & j=Q_{2}.
\end{cases}\end{split}
\label{eq:pi_0j_star}
\end{equation}
\end{lem}
\begin{IEEEproof}
From the proof of Theorem \ref{thm:opt_structure}, the optimal probability
$\pi_{(0,j)}^{*}$ is a function of $\pi_{(0,0)}^{*}$, as given by
(\ref{eq:pi_0j_star}).
\end{IEEEproof}
From (\ref{eq:pi_0j_star}), we get $\pi_{0}^{*}=\sum_{j=0}^{Q_{2}}\pi_{(0,j)}^{*}=\alpha\pi_{(0,0)}^{*}$,
where 
\[
\alpha=\sum_{i=0}^{Q_{2}-1}\phi^{-i}+\phi^{-(Q_{2}-1)}\phi_{1}^{-1}=\begin{cases}
(Q_{2}+\phi_{1}^{-1}), & \phi=1,\\
\frac{\phi_{1}\phi^{Q_{2}}+\phi-\phi_{1}-1}{\phi^{Q_{2}-1}(\phi-1)\phi_{1}}, & \phi\neq1.
\end{cases}
\]
From the results obtained in Theorem \ref{thm:opt_structure}, we
show that the optimal $\pi_{(i,0)}^{*}$ for all $i>0$ are functions
of $\pi_{(0,0)}^{*}$. Further, taking advantage of the dependance
of $\bar{P}$ on $\pi_{(0,0)}^{*}$, we can derive the closed-form
optimal solution $\pi_{(i,0)}^{*}$ in Corollary \ref{cor:opt_solution_SA}. 
\begin{cor}
\label{cor:opt_solution_SA}In Case I, when $p_{max}\geq\tilde{p}_{0}=\mu_{2}\alpha^{-1}$,
we have \textup{$\pi_{(0,0)}^{*}=\alpha^{-1}$} and $\pi_{(i,0)}^{*}=0$
for all $i>0$, respectively. \textup{\emph{When}}\textup{ }$\eta_{1}-\eta_{2}<p_{max}<\tilde{p}_{0}$,
$\pi_{(0,0)}^{*}=\frac{p_{max}-(\mu_{2}-\mu_{0})}{\mu_{2}-\alpha(\mu_{2}-\mu_{0})}$,
and \textup{$\pi_{(i,0)}^{*}$} $(i>0)$ is given by
\begin{equation}
\begin{split}\pi_{(i,0)}^{*} & =\begin{cases}
\pi_{(0,0)}^{*}\phi^{i}, & i\leq i^{*}-1,\\
1-\alpha\pi_{(0,0)}^{*}-\pi_{(0,0)}^{*}\sum_{i=1}^{i^{*}-1}\phi^{i}, & i=i^{*},\\
0, & i>i^{*},
\end{cases}\end{split}
\label{eq:opt_prob_pi_1i}
\end{equation}
where the optimal threshold $i^{*}$ is obtained as 
\begin{equation}
i^{*}=\Omega_{\phi}(\pi_{(0,0)}^{*},1-\alpha\pi_{(0,0)}^{*})\label{eq:opt_i_sa}
\end{equation}
with the function $\Omega_{\phi}(a,b)$ defined as
\[
\Omega_{\phi}(a,b):=\max_{a\sum\limits _{m=1}^{i-1}\phi^{m}\leq b}i=\begin{cases}
\lfloor\frac{b}{a}\rfloor+1, & \phi=1,\\
\lfloor\log_{\phi}\frac{(a+b)\phi-b}{a}\rfloor, & \phi\neq1.
\end{cases}
\]
\end{cor}
\begin{IEEEproof}
The proof is deferred to Appendix \ref{sub:Proof-of-Corollary8}.
\end{IEEEproof}
From Eqs. (\ref{eq:pi_0j_star}), (\ref{eq:opt_prob_pi_1i}) and (\ref{eq:opt_i_sa}),
one can see that the optimal steady-state probabilities $\pi_{(i,0)}^{*}$
and $\pi_{(0,j)}^{*}$, and the optimal threshold $i^{*}$ are solely
determined by the maximum average power $p_{max}$ for given $\eta_{1}$,
$\eta_{2}$ and $Q_{2}$. We also show that $\pi_{(i,0)}^{*}=0$ for
all $i>i^{*}$. This indicates that the length of the packet queue
never exceeds the threshold $i^{*}$. Hence, no packet loss will be
induced as long as the queue capacity $Q_{1}$ is larger than $i^{*}$.

\subsubsection{Case II and Case III}

In Case II with $k_{1}=1$ and $k_{2}>1$ and Case III with $k_{1}>1$
and $k_{2}=1$, it is challenging to derive a closed-form optimal
solution to the LP problem (\ref{eq:LP_problem}). Based on the result
in Theorem \ref{thm:opt_structure}, we then develop an algorithm
to find the optimal solutions for these two cases.

In Theorem \ref{thm:opt_structure}, we show that the optimal solution
corresponds to the threshold based transmission scheme. The optimal
threshold can be determined by comparing the power constraint $p_{max}$
to the power thresholds $\{\tilde{p}_{m}\}$ ($m\geq0$). In particular,
$\tilde{p}_{m}$ can be computed as 
\begin{equation}
\tilde{p}_{m}=\bm{\pi}_{m}^{'}\bm{a}_{0},
\end{equation}
where $\bm{\pi}_{m}^{'}$ is the solution to the following linear
equations
\begin{equation}
\begin{cases}
\bm{\pi}\bm{a}_{i}^{u} & =0\,(i=1,\cdots,m),\\
\bm{\pi}\bm{a}_{i}^{l} & =0\,(i=m+1,\cdots,Q_{1}),\\
\bm{\pi}\tilde{\bm{P}}_{s} & =\bm{0},\\
\bm{\pi}\bm{e} & =1.
\end{cases}\label{eq:opt_pi_m}
\end{equation}
Let $\bm{b}^{'}=[0,\cdots,0,1]$ be a $1\times N$ row vector, and
$\bm{A}_{m}^{'}=[\bm{a}_{1}^{u},\cdots,\bm{a}_{m}^{u},\bm{a}_{m+1}^{l},\cdots,\bm{a}_{Q_{1}}^{l},\tilde{\bm{P}}_{s},\bm{e}]$
be an $N\times N$ matrix. The solution to (\ref{eq:opt_pi_m}) can
be expressed as $\bm{\pi}_{m}^{'}=\bm{b}^{'}(\bm{A}_{m}^{'})^{-1}$.
Thus, the power threshold is equal to $\tilde{p}_{m}=\bm{\pi}_{m}^{'}\bm{a}_{0}=\bm{b}^{'}(\bm{A}_{m}^{'})^{-1}$. 

Once obtaining the power thresholds $\{\tilde{p}_{m}\}$, we can compute
the optimal solution $\bm{\pi}^{*}$ as follows.
\begin{cor}
\label{cor:opt_solution}The optimal solution to (\ref{eq:LP_problem})
for Cases II and III can be computed as
\begin{equation}
\bm{\pi}^{*}=\begin{cases}
\bm{\pi}_{0}^{'}, & \text{{\rm if } }p_{max}\geq\tilde{p}_{0},\\
\bm{b}\bm{A}^{-1}, & \text{{\rm if } }\tilde{p}_{i^{*}}\leq p_{max}<\tilde{p}_{i^{*}-1},
\end{cases}
\end{equation}
where $\bm{A}=[\bm{a}_{0},\bm{a}_{1}^{u},\cdots,\bm{a}_{i^{*}-1}^{u},\bm{a}_{i^{*}+1}^{l},\cdots,\bm{a}_{Q_{1}}^{l},\tilde{\bm{P}}_{s},\bm{e}]$
is an $N\times N$ matrix, and $\bm{b}=[p_{max},\cdots,0,1]$ is a
$1\times N$ row vector.\end{cor}
\begin{IEEEproof}
The proof is deferred to Appendix \ref{sub:Proof-of-Corollary9}.
\end{IEEEproof}
\emph{Remark}: By exploiting the structure of the optimal solution,
we can compute the optimal solution $\bm{\pi}^{*}$ by solving $(1+Q_{1})(1+Q_{2})$
independent linear equations. Based on the definition of the power
thresholds $\{\tilde{p}_{m}\}$, $\bm{\pi}^{*}$ can be alternatively
obtained by solving linear equations (\ref{eq:opt_pi_m}) when $p_{max}=\tilde{p}_{i^{*}}$,
$i.e.$, $\bm{\pi}^{*}=\bm{\pi}_{m}^{'}$. In Case II, since $\pi_{i}=\Theta_{l}(i,\tilde{\bm{\pi}}_{i-1})=0$,
we have $\pi_{(i,j)}^{*}=0$ for all $i>i^{*}$. And $\tilde{\bm{\pi}}_{i^{*}}^{*}$
can be obtained by solving $(1+i^{*})(1+Q_{2})$ linear equations:
$\tilde{\bm{\pi}}_{i^{*}}\bm{a}_{0}=p_{max}$, $\tilde{\bm{\pi}}_{i^{*}}\bm{a}_{i}^{u}=0\,(i=1,\cdots,i^{*}-1)$,
$\tilde{\bm{\pi}}_{i^{*}}\tilde{\bm{P}}_{i^{*}}=\bm{0}$ and $\tilde{\bm{\pi}}_{i^{*}}\bm{e}_{i^{*}}=1$.
\begin{table*}
\centering
\renewcommand{\arraystretch}{1.5} 

\caption{The optimal transmission parameters $g_{i}^{*}$ and $f_{i}^{*}$
for Cases I, II, and III.\label{Table_optimal_parameters}}

\begin{tabular}{|c|c|c|c|c|}
\hline 
\multicolumn{2}{|c|}{} & Case I with $k_{1}=k_{2}=1$ & Case II with $k_{1}=1$ and $k_{2}>1$ & Case III with $k_{1}>1$ and $k_{2}=1$\tabularnewline
\hline 
 & $0\leq i<i^{*}-k_{1}$ & \multicolumn{1}{c}{} & \multicolumn{1}{c}{0} & \tabularnewline
\cline{2-5} 
$g_{i}^{*}$ & $i=i^{*}-k_{1}$ & $1-\frac{1-\alpha\pi_{(0,0)}^{*}-\pi_{(0,0)}^{*}\sum_{i=1}^{i^{*}-1}\phi^{i}}{\pi_{(0,0)}^{*}\phi^{i^{*}}}$ & $1-\frac{\pi_{(i^{*},0)}^{*}\mu_{0}+\bar{\eta}_{1}\sum_{j=1}^{Q_{2}}\pi_{(i^{*},j)}^{*}}{\mu_{2}\pi_{(i^{*}-1,0)}^{*}}$ & $1-\frac{\bar{\eta}_{1}\pi_{i^{*}}^{*}-\pi_{(i^{*},0)}^{*}\mu_{1}-\eta_{1}\sum\limits _{m=i^{*}-k_{1}+1}^{i^{*}-1}\pi_{m}^{*}}{\mu_{2}\pi_{(i^{*}-k_{1},0)}^{*}}$\tabularnewline
\cline{2-5} 
 & $i>i^{*}-k_{1}$ & \multicolumn{1}{c}{} & \multicolumn{1}{c}{1} & \tabularnewline
\hline 
\hline 
\multicolumn{2}{|c|}{} & Case I and Case II & \multicolumn{2}{c|}{Case III with $k_{1}>1$ and $k_{2}=1$}\tabularnewline
\hline 
 & $0<i<i^{*}$ & \multirow{4}{*}{0} & \multicolumn{2}{c|}{0}\tabularnewline
\cline{2-2} \cline{4-5} 
\multirow{2}{*}{$f_{i}^{*}$} & \multirow{2}{*}{$i=i^{*}$} &  & \multicolumn{2}{c|}{0 $(i^{*}\geq k_{1})$}\tabularnewline
\cline{4-5} 
 &  &  & \multicolumn{2}{c|}{$\frac{\eta_{1}\sum\limits _{m=0}^{i^{*}-1}\pi_{m}^{*}-\bar{\eta}_{1}\pi_{i^{*}}^{*}+\pi_{(i^{*},0)}^{*}\mu_{1}}{\pi_{(i^{*},0)}^{*}\mu_{1}}$
$(i^{*}<k_{1})$}\tabularnewline
\cline{1-2} \cline{4-5} 
 & $i>i^{*}$ &  & \multicolumn{2}{c|}{1}\tabularnewline
\hline 
\end{tabular}
\end{table*}

\begin{algorithm}[t]
\caption{Finding the optimal solution for Cases II and III. }

\begin{algorithmic}[1]

\STATE Initialization: set $Q_{1}$ to be a large constant. 

\IF{ $p_{max}\leq k_{1}\eta_{1}-k_{2}\eta_{2}$}

\STATE The optimal solution and parameters do not exist.

\ELSE 

\STATE Compute $\bm{\pi}_{0}^{'}=\bm{b}^{'}(\bm{A}_{0}^{'})^{-1}$
and $\tilde{p}_{0}=\bm{\pi}_{0}^{'}\bm{a}_{0}$.

\IF{ $p_{max}\geq\tilde{p}_{0}$}

\STATE Set $\bm{\pi}^{*}=\bm{\pi}_{0}^{'}$. 

\ELSE

\STATE Set $m=1$.

\REPEAT

\STATE Compute $\bm{\pi}_{m}^{'}=\bm{b}^{'}(\bm{A}_{m}^{'})^{-1}$
and $\tilde{p}_{m}=\bm{\pi}_{m}^{'}\bm{a}_{0}$.

\IF{ $\tilde{p}_{m}\leq p_{max}<\tilde{p}_{m-1}$}

\STATE Set $i^{*}=m$. Compute $\bm{\pi}^{*}=\bm{b}\bm{A}^{-1}$.
Exit.

\ENDIF

\UNTIL{$m>Q_{1}$}

\STATE Set $\bm{\pi}^{*}=\bm{\pi}_{Q_{1}}^{'}$, and set $i^{*}=\infty$.

\ENDIF

\ENDIF

\end{algorithmic}
\end{algorithm}
 Based on the above discussion, we develop an algorithm, $i.e.$,
Algorithm 1, to show how to find the optimal solution $\pi_{(i,j)}^{*}$
to the LP problem for Case II and Case III. The optimal threshold
$i^{*}$ is sought iteratively by comparing the maximum allowable
power consumption $p_{max}$ with the power thresholds $\{\tilde{p}_{m}\}$.
Once locating the threshold $i^{*}$, the optimal steady-state probabilities
$\pi_{(i,j)}^{*}$ can be obtained by solving an LP problem. There
are two exceptions: (1) when $p_{max}\leq k_{1}\eta_{1}-k_{2}\eta_{2}$,
the queueing system is not stable and the optimal solution does not
exist; and (2) when the iteration number exceeds the sufficiently
large data queue length $Q_{1}$, we regard the optimal threshold
$i^{*}$ as the infinity. Given $Q_{1}$, the algorithm runs at most
$Q_{1}$ iterations, and in each iteration the computation complexity
of solving $N$ linear equations is $\mathcal{O}(N^{3})$. Hence,
the computation complexity of this algorithm can be roughly estimated
as $\mathcal{O}(Q_{1}(1+Q_{1})^{3}(1+Q_{2})^{3})$. For a relatively
small $Q_{2}$, the complexity can be approximated as $\mathcal{O}(Q_{1}^{4})$. 

By comparison of Case I and Case II, we notice that changing the
number of energy packets arriving each time, $k_{2}$, does not change
the property of the optimal results. From this perspective, it is
feasible to deal with the case when $k_{1}>1$ and $k_{2}>1$ using
the same method as in Case III. Firstly, we formulate a concrete Markov
chain for a pair of such $k_{1}$ and $k_{2}$ and find the mutual
relations between the states. Secondly, we construct an LP problem
under the power consumption constraint, which manifests as the constant
linear combination of the steady-state probabilities. Finally, we
can adopt Algorithm 1 to find the optimal solution and the optimal
transmission parameters.

\subsection{The Optimal Transmission Parameters}

\noindent By exploiting the local equilibrium equations and the corresponding
optimal solution $\bm{\pi}^{*}$, we then obtain the optimal transmission
parameters $\{g_{i}^{*}\}$ and $\{f_{i}^{*}\}$ for Cases I, II and
III. 
\begin{cor}
\label{cor:opt_gi_fi}When $p_{max}\geq\tilde{p}_{0}$, the optimal
transmission parameters are given by $g_{i}^{*}=1$ $(i\geq0)$ and
$f_{i}^{*}=0$ $(i>0)$; When $k_{1}\eta_{1}-k_{2}\eta_{2}<p_{max}<\tilde{p}_{0}$,
the optimal transmission parameters $\{g_{i}^{*}\}$ and $\{f_{i}^{*}\}$
are listed in Table \ref{Table_optimal_parameters}. \end{cor}
\begin{IEEEproof}
The proof is deferred to Appendix \ref{sub:Proof-of-Corollary10}.
\end{IEEEproof}
\emph{Remark}: Note that $p_{max}\ge\tilde{p}_{0}$ indicates that
the allowable backup energy supply is sufficient so that the source
can use the reliable energy whenever it needs. In this scenario, packet
delivery is guaranteed in each slot, and there will be no backlogged
packets in Case I and Case II, while in Case III, the data queue may
still accumulate as each data arrival brings in multiple packets while
at most one packet is delivered in each slot. When $k_{1}\eta_{1}-k_{2}\eta_{2}<p_{max}<\tilde{p}_{0}$,
the source should transmit according to the optimal threshold $i^{*}$.
For Case I and Case II, the utilization of power from the RES could
happen only in two scenarios: when a new packet arrives but no harvested
energy can be used, and the data packet queue length is equal to $i^{*}-1$
and $i^{*}$, respectively. In the former case, the source transmits
using the power from the RES with the probability $g_{i^{*}-1}^{*}<1$,
while in the latter case, the source will transmit using the energy
from the RES definitely with $g_{i^{*}}^{*}=1$. In Case III, the
source should transmit using the reliable energy as soon as the data
queue length exceeds the threshold $i^{*}$, no matter whether there
is a new data arrival. That is, $g_{i-k_{1}}^{*}=1$ and $f_{i}^{*}=1$
are set for all $i>i^{*}$. Once getting the optimal $g_{i}^{*}$
and $f_{i}^{*}$, we can compute the optimal steady-state probabilities
$\pi_{(i,j)}^{*}$ and the corresponding minimum average delay $\bar{D}^{*}=\frac{1}{k_{1}\eta_{1}}\sum_{i=1}^{Q_{1}}i\sum_{j=0}^{Q_{2}}\pi_{(i,j)}^{*}$,
which depends on the allowable reliable energy $p_{max}$. Hence,
$\bar{D}^{*}$ is an implicit function of $p_{max}$. As shown by
simulation results in the next section, the average queuing delay
monotonically decreases with the increase of the power $p_{max}$.

\section{Simulation Results\label{sec:Simulation-Results}}

\noindent In this section, simulation results are presented to demonstrate
the performance of the proposed scheduling scheme and validate our
theoretical analysis.

In simulations, the packet and energy arrival processes are modeled
by generating two Bernoulli random variables with the parameters $\eta_{1}$
and $\eta_{2}$, respectively, at the beginning of each time slot.
The packet transmissions are scheduled according to our proposed policy.
And we apply the optimal transmission parameters $f_{i}^{*}$ and
$g_{i}^{*}$ listed in Table \ref{Table_optimal_parameters} to get
the optimal delay-power tradeoff curves. Each simulation runs over
$10^{6}$ time slots. In the figures, the lines and the marks 'o'
indicate theoretical and simulation results, respectively. One can
see that theoretical and simulation results match well. 

\begin{figure}[t]
\centering
\renewcommand{\figurename}{Fig.}\subfigure[$\eta_1=\eta_2=0.3$]{\label{fig:delay_power_sa}\includegraphics[width=0.9\columnwidth]{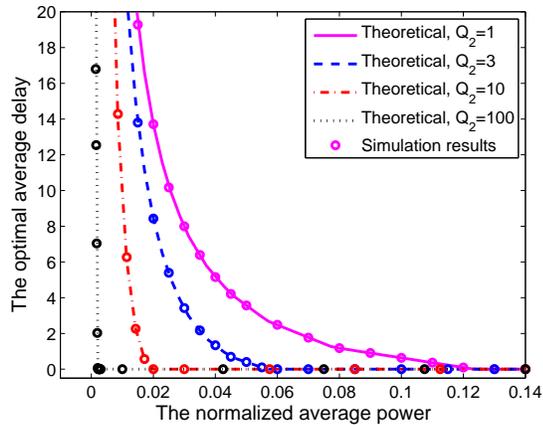}}\\\centering\subfigure[$\eta_1=0.5, Q_2=1$]{\label{fig:delay_power_sa_lambda2}\includegraphics[width=0.9\columnwidth]{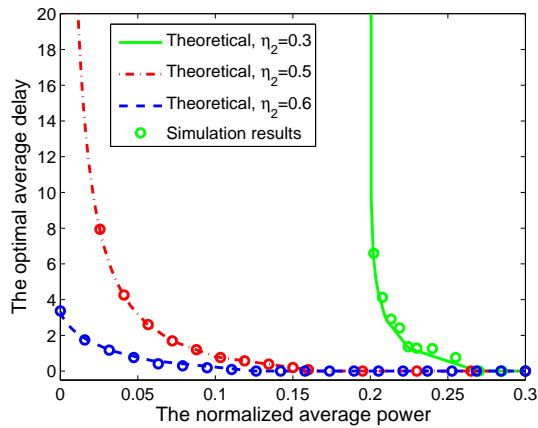}}

\caption{The delay-power curve for Case I. }
\label{fig:performance_CaseI}
\end{figure}

Fig.$\,$\ref{fig:performance_CaseI} plots the optimal delay-power
tradeoff performance for Case I. It is observed from Fig.$\,$\ref{fig:delay_power_sa}
that the minimum average queueing delay monotonically decreases with
the increase of the maximum power consumption $p_{max}$, which contributes
to the growth of the service rate. That is, when more power can be
drawn from the RES, the packets will be transmitted more quickly and
the queueing delay is reduced. One can see that the minimum average
queueing delay decreases from infinity to zero, when $p_{max}$ grows
from zero to $\tilde{p}_{0}$, which is equal to $\frac{\mu_{2}}{Q_{2}+\phi_{1}^{-1}}$
in the case of $\eta_{1}=\eta_{2}$. Hence, the decreasing rate grows
with the increase of the battery capacity $Q_{2}$. This means that
a larger $Q_{2}$ leads to a much smaller queueing delay, since less
harvested energy is wasted due to the limitation of the battery capacity.
Fig.$\,$\ref{fig:delay_power_sa_lambda2} demonstrates the optimal
delay-power performance for different energy arrival rates $\eta_{2}$.
When $\eta_{2}\leq\eta_{1}$, the average delay is infinite at $p_{max}=0$,
since the arrival rate is greater than or equal to the service rate.
Therefore, the source should exploit extra energy from the RES to
transmit backlogged packets, corresponding to a positive $p_{max}$.
While the source can rely only on the harvested energy to transmit,
$i.e.$, $p_{max}=0$, when $\eta_{2}>\eta_{1}$. 

\begin{figure}[t]
\centering
\renewcommand{\figurename}{Fig.}\includegraphics[width=0.9\columnwidth]{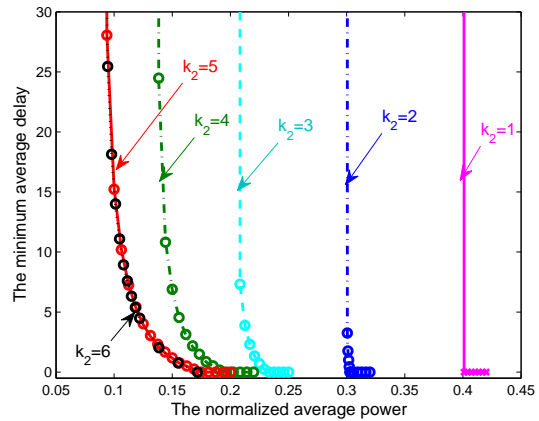}

\caption{The optimal delay-power performance for Case II with $\eta_{1}=0.5$,
$\eta_{2}=0.1$ and $Q_{2}=5$. }
\label{fig:performance_CaseII}
\end{figure}

Fig.$\,$\ref{fig:performance_CaseII} shows the optimal delay-power
tradeoff performance of the proposed scheme in Case II with $k_{2}=2,\cdots,6$.
The optimal delay-power curve of the case with $k_{1}=k_{2}=1$ is
also plotted for comparison. In this experiment, we set $\eta_{1}=0.5$,
$\eta_{2}=0.1$ and $Q_{2}=5$. From this figure, one can see that
there exists an optimal delay-power tradeoff for each $k_{2}$. The
average queueing delay monotonically decreases with the increase of
the maximum allowable power consumption $p_{max}$ from the RES due
to the enhanced service rate. For the same reason, a larger $k_{2}$
means a higher amount of energy harvested each time, and leads to
a much better delay-power tradeoff. It is also observed the delay-power
curves of $k_{2}=5$ and $k_{2}=6$ are almost identical to each other.
This owes to the fact that in the case of $k_{2}=6$, a part of harvest
energy is wasted when recharging the battery with capacity $Q_{2}=5$.

\begin{figure}[t]
\centering
\renewcommand{\figurename}{Fig.}\includegraphics[width=0.9\columnwidth]{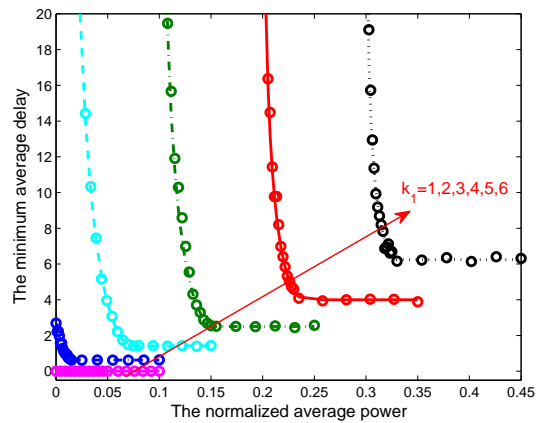}

\caption{The delay-power performance for Case III with $\eta_{1}=0.1,\eta_{2}=0.3$,
and $Q_{2}=5$. }
\label{fig:performance_CaseIII}
\end{figure}

Similarly, we plot the optimal delay-power curves of the proposed
scheme for Case III with different $k_{1}$ in Fig.$\,$\ref{fig:performance_CaseIII}.
We set $\eta_{1}=0.1,\eta_{2}=0.3$, and $Q_{2}=5$. Similar to Case
II shown in Fig.$\,$\ref{fig:performance_CaseII}, a higher $p_{max}$
induces reduced average queueing delay thanks to the enhanced service
rate. The only difference between them is the behavior of the minimum
average delay $\bar{D}^{*}$. In Case I with $k_{1}=k_{2}=1$, the
average queueing delay is equal to zero if there exists sufficient
energy whether from the battery or the RES, since one newly arriving
data packet can always be delivered immediately. In Case III, however,
at most one of $k_{1}$ data packets that newly arrive at this slot
can be delivered, and the other packets shall wait for the next transmission
opportunity. And more packets are queued when the data arrival rate
is increased due to the growth of $k_{1}$ or $\eta_{1}$. As shown
in Fig.$\,$\ref{fig:performance_CaseIII}, $\bar{D}^{*}$ increases
with the increase of $k_{1}$.

\section{Conclusions\label{sec:Conclusions}}

\noindent In this paper, we investigated the delay optimal scheduling
problem over a communication link. The source node can rely on energy
supply either from an energy harvesting battery of finite capacity
or from the RES subject to a maximum power consumption from the RES.
Using the two-dimensional Markov chain modeling, we formulated an
LP problem and studied the structure of the optimal solution. As a
result, we obtained the optimal scheduling policy through rigorous
derivation and algorithm design. 

It is found that the source should schedule packet transmissions according
to a critical threshold on the data queue length. Specifically, the
source should always wait for the harvested energy when the data queue
length is below the optimal threshold $i^{*}$, and resort to the
RES when the data queue length exceeds the threshold $i^{*}$ while
no harvested energy can be exploited. The optimal threshold $i^{*}$
is determined by the maximum allowable power from the RES $p_{max}$.
 Simulation results confirmed our theoretical analysis. It was shown
that there always exists an optimal delay-power tradeoff and its decreasing
rate depends on the energy arrival rate and the battery capacity. 

In this work, we assume that the Bernoulli data and energy arrival
processes generate integral packets probabilistically, and only one
data packet is transmitted in each slot. In the future, we will extend
the study to the scenario where rate-flexible physical-layer transmissions
are scheduled based on the randomly available amount of harvested
energy and time-varying wireless channel conditions.

\appendix

\subsection{Proof of Lemma \ref{lem:ave_power}\label{sub:Proof-of-Lemma2}}

\noindent By applying the stochastic scheduling scheme described
in Section \ref{sec:System-Model}.2, the source shall resort to the
RES only when the harvested energy is not available in the current
slot. Thus, the state set $\mathcal{Q}_{p}$ (c.f. (\ref{eq:av_power}))
is given by $\{(0,0),(1,0),\cdots,(Q_{1},0)\}$. Recall that the source
will draw the reliable energy to transmit with probability $g_{i}$
if new data packets arrive and with probability $f_{i}$ if no data
packets arrive, respectively. Hence, the reliable energy consumption
at state $(i,0)$ can be expressed as $\omega_{(0,0)}(p)=\mu_{2}g_{0}$
and $\omega_{(i,0)}(p)=\mu_{2}g_{i}+\mu_{1}f_{i}$ $(i>0)$, respectively.
Consequently, the normalized average power consumed from the RES can
be obtained as $\bar{P}=\sum_{i=0}^{Q_{1}}\pi_{(i,0)}\omega_{(i,0)}(p)=\sum_{i=0}^{Q_{1}}\pi_{(i,0)}\mu_{2}g_{i}+\sum_{i=1}^{Q_{1}}\pi_{(i,0)}\mu_{1}f_{i}$.
The following result eliminates the dependence of $\bar{P}$ on the
transmission parameters and presents a unified expression for all
three cases.

1) In Case I, from Fig.$\,$\ref{fig:markov_chain_onefig}(c), the
local equilibrium equation of the Markov chain can be expressed as
\begin{equation}
\pi_{(i,0)}\lambda_{1,i}=\pi_{(i+1,0)}\tilde{\mu}_{1,i+1},\,\pi_{(0,j)}\mu_{0}=\pi_{(0,j+1)}\mu_{2},\label{eq:local_equilibrium}
\end{equation}
for all $i\in\{0,\cdots,Q_{1}-1\}$ and $j\in\{0,\cdots,Q_{2}-2\}$,
and $\pi_{(0,Q_{2}-1)}\mu_{0}=\pi_{(0,Q_{2})}\eta_{1}$.  From (\ref{eq:local_equilibrium}),
we have $\pi_{(0,0)}\lambda_{1,0}=\sum_{i=1}^{Q_{1}}\pi_{(i,0)}(\tilde{\mu}_{1,i}-\lambda_{1,i})$,
where $\pi_{(Q_{1},0)}\lambda_{1,Q_{1}}=0$ is introduced for notational
convenience. With $\lambda_{1,i}=\mu_{2}(1-g_{i})$ (c.f. (\ref{eq:par_lambda1i_mu_1i_general}))
and $\tilde{\mu}_{1,i}=\mu_{0}+\mu_{1}f_{i}$ (c.f. (\ref{eq:par_mu_1i_case1})),
the normalized average power consumption from the RES $\bar{P}$ can
also be expressed as 
\[
\begin{split}\bar{P}= & \pi_{(0,0)}\mu_{2}g_{0}+\sum\nolimits _{i=1}^{Q_{1}}\pi_{(i,0)}(\mu_{2}g_{i}+\mu_{1}f_{i})\\
= & \pi_{(0,0)}(\mu_{2}-\lambda_{1,0})+\sum\nolimits _{i=1}^{Q_{1}}\pi_{(i,0)}(\mu_{2}-\lambda_{1,i}+\tilde{\mu}_{1,i}-\mu_{0})\\
= & \pi_{(0,0)}\mu_{2}+(\mu_{2}-\mu_{0})\sum\nolimits _{i=1}^{Q_{1}}\pi_{(i,0)}.
\end{split}
\]
Hence, we get $\xi_{0}=\mu_{2}$, $\xi_{i}=\mu_{2}-\mu_{0}$ for all
$i\in\mathcal{Q}_{1}^{R}$ and $\zeta_{i}=0$ for all $i\in\mathcal{Q}_{1}$. 

2) From Fig.$\,$\ref{fig:markov_chain_onefig}(d) for Case II, the
 local equilibrium equation at state $(0,0)$ can be expressed as
$\pi_{(1,0)}\mu_{1,1}-\pi_{(0,0)}\lambda_{1,0}=\eta_{2}\pi_{(0,0)}-\mu_{2}\pi_{(0,1)}-\mu_{1}\pi_{(1,1)}$.
Following a similar procedure, the  local equilibrium equation at
state $(i,0)$ ($i\geq1$) is thus
\[
\begin{split} & \pi_{(i+1,0)}\mu_{1,i+1}-\pi_{(i,0)}\lambda_{1,i}=\pi_{(i,0)}\mu_{1,i}-\pi_{(i-1,0)}\lambda_{1,i-1}\\
 & +(\mu_{0}+\mu_{3})\pi_{(i,0)}-\mu_{2}\pi_{(i,1)}-\mu_{1}\pi_{(i+1,1)}\\
= & \eta_{2}\sum\nolimits _{m=0}^{i}\pi_{(m,0)}-\mu_{2}\sum\nolimits _{m=0}^{i}\pi_{(m,1)}-\mu_{1}\sum\nolimits _{m=1}^{i+1}\pi_{(m,1)},
\end{split}
\]
where the second equality is obtained through recursion of $(\pi_{(i,0)}\mu_{1,i}-\pi_{(i-1,0)}\lambda_{1,i-1})$.
Hence, we can compute the normalized average power consumption as
\[
\begin{split}\bar{P} & =\sum\nolimits _{i=0}^{Q_{1}}\pi_{(i,0)}(\mu_{2}-\lambda_{1,i})+\sum\nolimits _{i=1}^{Q_{1}}\pi_{(i,0)}\mu_{1}f_{i}\\
= & \mu_{2}\sum\nolimits _{i=0}^{Q_{1}}\pi_{(i,0)}+\sum\nolimits _{i=1}^{Q_{1}}\left(\pi_{(i,0)}\mu_{1,i}-\pi_{(i-1,0)}\lambda_{1,i-1}\right)\\
= & \sum\nolimits _{i=0}^{Q_{1}}\pi_{(i,0)}(\mu_{2}+\eta_{2}(Q_{1}-i))\\
 & -\mu_{2}Q_{1}\pi_{(0,1)}-\sum\nolimits _{i=1}^{Q_{1}}(\bar{\eta}_{2}(Q_{1}-i)+\mu_{1})\pi_{(i,1)},
\end{split}
\]
where the second equality is due to the fact that $\mu_{1,i}=\mu_{1}f_{i}$,
and the third equality is obtained through the summation of $\pi_{(i,0)}\mu_{1,i}-\pi_{(i-1,0)}\lambda_{1,i-1}$
over all $i>0$. In this way, we get $\xi_{i}=\mu_{2}+\eta_{2}(Q_{1}-i)$
for all $i\in\mathcal{Q}_{1}$, $\zeta_{0}=\mu_{2}Q_{1}$, and $\zeta_{i}=\bar{\eta}_{2}(Q_{1}-i)+\mu_{1}$
for all $i\in\mathcal{Q}_{1}^{R}$. 

3) In Case III, the corresponding Markov chain is depicted in Fig.$\,$\ref{fig:markov_chain_onefig}(e).
When $i<k_{1}-1$, the local balance equation at state $(i,0)$ is
given by 
\[
\begin{split} & \tilde{\mu}_{1,i+1}\pi_{(i+1,0)}=\pi_{(i,0)}\tilde{\mu}_{1,i}+\pi_{(i,0)}(\lambda_{1,i}+\tilde{\lambda}_{1,i})-\mu_{1}\pi_{(i+1,1)}\\
= & \pi_{(0,0)}\mu_{0}+\sum\nolimits _{m=0}^{i}\pi_{(m,0)}\eta_{1}-\sum\nolimits _{m=1}^{i+1}\mu_{1}\pi_{(m,1)},
\end{split}
\]
where $\lambda_{1,i}+\tilde{\lambda}_{1,i}=\eta_{1}$ (c.f. (\ref{eq:par_lambda_case3}))
is applied. When $i\geq k_{1}-1$, the local balance equation at state
$(i,0)$ can be expressed as \textbf{
\[
\begin{split} & \pi_{(i-k_{1}+1,0)}\tilde{\lambda}_{1,i-k_{1}+1}+\tilde{\mu}_{1,i+1}\pi_{(i+1,0)}\\
= & \pi_{(i,0)}\tilde{\mu}_{1,i}-\lambda_{1,i-k_{1}}\pi_{(i-k_{1},0)}+\pi_{(i,0)}(\lambda_{1,i}+\tilde{\lambda}_{1,i})\\
 & -\mu_{2}\pi_{(i-k_{1}+1,1)}-\mu_{1}\pi_{(i+1,1)}=\eta_{1}\sum_{m=i-k_{1}+1}^{i}\pi_{(m,0)}\\
 & +\pi_{(0,0)}\mu_{0}-\mu_{1}\sum\nolimits _{m=1}^{i+1}\pi_{(m,1)}-\mu_{2}\sum\nolimits _{m=0}^{i-k_{1}+1}\pi_{(m,1)}.
\end{split}
\]
}Then, we can compute the normalized average power consumption as
\[
\begin{split}\bar{P}= & \sum\nolimits _{i=0}^{Q_{1}}\pi_{(i,0)}\mu_{2}g_{i}+\sum\nolimits _{i=1}^{Q_{1}}\pi_{(i,0)}\mu_{1}f_{i}\\
= & \sum\nolimits _{i=k_{1}-1}^{Q_{1}-1}(\pi_{(i-k_{1}+1,0)}\tilde{\lambda}_{1,i-k_{1}+1}+\tilde{\mu}_{1,i+1}\pi_{(i+1,0)})\\
 & +\sum\nolimits _{i=0}^{k_{1}-2}\tilde{\mu}_{1,i+1}\pi_{(i+1,0)}-\mu_{3}\pi_{(0,0)}-\eta_{2}\sum\nolimits _{i=1}^{Q_{1}}\pi_{(i,0)}\\
= & \sum\nolimits _{i=0}^{Q_{1}}\xi_{i}\pi_{(i,0)}-\sum\nolimits _{i=0}^{Q_{1}}\zeta_{i}\pi_{(i,1)}.
\end{split}
\]
where the second equality is due to the fact that $\tilde{\lambda}_{1,i}=\mu_{3}+\mu_{2}g_{i}$
(c.f. (\ref{eq:par_lambda_case3})) and $\tilde{\mu}_{1,i}=\mu_{0}+\mu_{1}f_{i}$
(c.f. (\ref{eq:par_mu_1i_case1})), the last equality is obtained
via the summation of $\pi_{(i-k_{1}+1,0)}\tilde{\lambda}_{1,i-k_{1}+1}+\tilde{\mu}_{1,i+1}\pi_{(i+1,0)}$
and $\tilde{\mu}_{1,i+1}\pi_{(i+1,0)}$ over $i\geq k_{1}-1$ and
$i<k_{1}-1$, respectively. As a result, we can compute the coefficients
$\xi_{i}$ and $\zeta_{i}$ as listed in Table \ref{Table_coefficients}.

\subsection{Proof of Lemma \ref{lem:relation_pi}\label{sub:Proof-of-Lemma3}}

\noindent We will prove that for each $i$, the probability $\pi_{i}$
satisfies the inequality (\ref{eq:bounds_pi}) for Cases I, II and
III, respectively. 

1) In Case I, we have $\lambda_{1,i}=\mu_{2}(1-g_{i})$ and $\tilde{\mu}_{1,i}=\mu_{0}+\mu_{1}f_{i}$,
which satisfy $0\leq\lambda_{1,i}\leq\mu_{2},\,\mu_{0}\leq\tilde{\mu}_{1,i}\leq\mu_{0}+\mu_{1}=\bar{\eta}_{1}$,
since $0\leq g_{i}\leq1$ and $0\leq f_{i}\leq1$. From the local
equilibrium equation (\ref{eq:local_equilibrium}), we have $0\leq\pi_{i}=\pi_{(i,0)}\leq\phi\pi_{(i-1,0)}$,
since $0\leq\frac{\pi_{(i,0)}}{\pi_{(i-1,0)}}=\frac{\lambda_{1,i-1}}{\tilde{\mu}_{1,i}}\leq\frac{\mu_{2}}{\mu_{0}}=\phi$.
In this case, we have $\Theta_{u}(i,\tilde{\bm{\pi}}_{i})=\phi\pi_{(i-1,0)}$
when $g_{i-1}=0$ and $f_{i}=0$, and $\Theta_{l}(i,\tilde{\bm{\pi}}_{i-1})=0$
when $g_{i-1}=1$. 

2) From Fig.$\,$\ref{fig:markov_chain_onefig}(d) in Case II, the
local balance equation $\pi_{(i-1,0)}\lambda_{1,i-1}=\pi_{(i,0)}(\mu_{1,i}+\mu_{0})+(\mu_{0}+\mu_{1})\sum_{j=1}^{Q_{2}-1}\pi_{(i,j)}+\bar{\eta}_{1}\pi_{(i,Q_{2})}$
holds for all $i>0$, thus leading to $\pi_{(i-1,0)}\lambda_{1,i-1}=\pi_{(i,0)}(\mu_{1,i}+\mu_{0})+\bar{\eta}_{1}\sum\nolimits _{j=1}^{Q_{2}}\pi_{(i,j)}$
because of $\mu_{0}+\mu_{1}=\bar{\eta}_{1}$. Hence, we have $\bar{\eta}_{1}\pi_{i}=\bar{\eta}_{1}\sum\nolimits _{j=0}^{Q_{2}}\pi_{(i,j)}=\pi_{(i-1,0)}\lambda_{1,i-1}-\pi_{(i,0)}(\mu_{1,i}+\mu_{0})+\bar{\eta}_{1}\pi_{(i,0)}$.
Since $0\leq g_{i}\leq1$ and $0\leq f_{i}\leq1$, we get $0\leq\lambda_{1,i}=\mu_{2}(1-g_{i})\leq\mu_{2}$
and $\mu_{0}\leq\mu_{1,i}+\mu_{0}=\mu_{1}f_{i}+\mu_{0}\leq\bar{\eta}_{1}$.
With the varying parameter $\{\lambda_{1,i}\}$ and $\{\mu_{1,i}\}$,
we further have $0\leq\pi_{i}\leq\Theta_{u}(i,\tilde{\bm{\pi}}_{i})$,
where $\pi_{i}=\Theta_{u}(i,\tilde{\bm{\pi}}_{i})=\tau\bar{\eta}_{2}\pi_{(i-1,0)}+\pi_{(i,0)}\bar{\eta}_{2}$
when $\lambda_{1,i-1}=\mu_{2}$ and $\mu_{1,i}+\mu_{0}=\mu_{0}$ ($g_{i-1}=f_{i}=0$),
and $\pi_{i}=\Theta_{l}(i,\tilde{\bm{\pi}}_{i-1})=0$ when $\lambda_{1,i-1}=0$
and $\mu_{1,i}+\mu_{0}=\bar{\eta}_{1}$ ($g_{i-1}=f_{i}=1$), respectively.

3) In Case III, from Fig.$\,$\ref{fig:markov_chain_onefig}(e), the
 local equilibrium equation between states $(i-1,j)$ and $(i,j)$
$(j\in\mathcal{Q}_{2})$ can be expressed as $\sum_{m=0}^{i-1}\pi_{(m,0)}(\lambda_{1,m}+\tilde{\lambda}_{1,m})+(\mu_{2}+\mu_{3})\sum_{m=0}^{i-1}\sum_{j=1}^{Q_{2}-1}\pi_{(m,j)}+\eta_{1}\sum_{m=0}^{i-1}\pi_{(m,Q_{2})}=\pi_{(i,0)}\tilde{\mu}_{1,i}+(\mu_{0}+\mu_{1})\sum_{j=1}^{Q_{2}-1}\pi_{(i,j)}+\bar{\eta}_{1}\pi_{(i,Q_{2})}$
for all $i<k_{1}$. With $\lambda_{1,i-1}+\tilde{\lambda}_{1,i-1}=\mu_{2}+\mu_{3}=\eta_{1}$
and $\mu_{0}+\mu_{1}=\bar{\eta}_{1}$, it can be rewritten as $\eta_{1}\sum_{m=0}^{i-1}\pi_{m}=\pi_{(i,0)}\tilde{\mu}_{1,i}+\bar{\eta}_{1}\sum_{j=1}^{Q_{2}}\pi_{(i,j)}$. 

Since $\mu_{0}\leq\tilde{\mu}_{1,i}=\mu_{0}+\mu_{1}f_{i}\leq\bar{\eta}_{1}$,
we get $\Theta_{l}(i,\tilde{\bm{\pi}}_{i-1})\leq\bar{\eta}_{1}\sum_{j=0}^{Q_{2}}\pi_{(i,j)}\leq\Theta_{u}(i,\tilde{\bm{\pi}}_{i})$.
Thus, $\pi_{i}=\Theta_{u}(i,\tilde{\bm{\pi}}_{i})=\frac{\mu_{1}}{\bar{\eta}_{1}}\pi_{(i,0)}+\frac{\eta_{1}}{\bar{\eta}_{1}}\sum_{m=0}^{i-1}\pi_{m}$
when $\tilde{\mu}_{1,i}=\mu_{0}$ ($f_{i}=0$), and $\pi_{i}=\Theta_{l}(i,\tilde{\bm{\pi}}_{i-1})=\frac{\eta_{1}}{\bar{\eta}_{1}}\sum_{m=0}^{i-1}\pi_{m}$
when $\tilde{\mu}_{1,i}=\bar{\eta}_{1}$ ($f_{i}=0$), respectively. 

Similarly, when $i\geq k_{1}$, the corresponding local equilibrium
equation between states $(i-1,j)$ and $(i,j)$ is given by $\eta_{1}\sum_{m=i-k_{1}+1}^{i-1}\pi_{m}=\pi_{(i,0)}\tilde{\mu}_{1,i}-\pi_{(i-k_{1},0)}\lambda_{1,i-k_{1}}+\bar{\eta}_{1}\sum_{j=1}^{Q_{2}}\pi_{(i,j)}$.
Since $0\leq\lambda_{1,i}=\mu_{2}(1-g_{i})\leq\mu_{2}$ and $\mu_{0}\leq\tilde{\mu}_{1,i}=\mu_{0}+\mu_{1}f_{i}\leq\bar{\eta}_{1}$,
we get $\eta_{1}\sum_{m=i-k_{1}+1}^{i-1}\pi_{m}\leq\bar{\eta}_{1}\sum_{j=0}^{Q_{2}}\pi_{(i,j)}\leq\mu_{2}\pi_{(i-k_{1},0)}+\mu_{1}\pi_{(i,0)}+\eta_{1}\sum_{m=i-k_{1}+1}^{i-1}\pi_{m}$.
Thus, $\pi_{i}=\Theta_{u}(i,\tilde{\bm{\pi}}_{i})=\frac{\mu_{2}}{\bar{\eta}_{1}}\pi_{(i-k_{1},0)}+\frac{\mu_{1}}{\bar{\eta}_{1}}\pi_{(i,0)}+\frac{\eta_{1}}{\bar{\eta}_{1}}\sum_{m=i-k_{1}+1}^{i-1}\pi_{m}$
when $\lambda_{1,i-k_{1}}=\mu_{2}$ and $\tilde{\mu}_{1,i}=\mu_{0}$
($g_{i-k_{1}}=0,f_{i}=0$), $\pi_{i}=\Theta_{l}(i,\tilde{\bm{\pi}}_{i-1})=\frac{\eta_{1}}{\bar{\eta}_{1}}\sum_{m=0}^{i-1}\pi_{m}$
when $\lambda_{1,i-k_{1}}=0$ and $\tilde{\mu}_{1,i}=\bar{\eta}_{1}$
($g_{i-k_{1}}=1,f_{i}=1$), respectively. 

Combining the above two cases: $i<k_{1}$ and $i\geq k_{1}$, we get
$\Theta_{u}(i,\tilde{\bm{\pi}}_{i})$ and $\Theta_{l}(i,\tilde{\bm{\pi}}_{i-1})$
as listed in Table \ref{Table_upper_lower_bounds}.

\subsection{Proof of Theorem \ref{thm:opt_threshold}\label{sub:Proof-of-Theorem4}}

\noindent Subject to the constraint (\ref{eq:LP_problem}.b), we
have $\bar{D}=\frac{1}{k_{1}\eta_{1}}\sum\nolimits _{i=1}^{Q_{1}}i\pi_{i}\geq\frac{1}{k_{1}\eta_{1}}\sum\nolimits _{i=1}^{Q_{1}}i\Theta_{l}(i,\tilde{\bm{\pi}}_{i-1})$.
This means that the average queuing delay $\bar{D}$, as a weighted
summation of $\pi_{i}$, can be minimized, if each $\pi_{i}$ chooses
its lower bound $\Theta_{l}(i,\tilde{\bm{\pi}}_{i-1})$ for all $i\geq1$.
From Lemma \ref{lem:relation_pi}, this happens when all the transmission
parameters satisfy $g_{i-k_{1}}=f_{i}=1$ for all $i>0$. This corresponds
to the scheduling policy based on the optimal threshold $i^{*}=0$
and the corresponding average power consumption from the RES is $\tilde{p}_{0}$.
Hence, when $p_{max}\geq\tilde{p}_{0}$, the minimum average queueing
delay $\bar{D}^{*}$ is obtained if $\pi_{i}^{*}=\Theta_{l}(i,\tilde{\bm{\pi}}_{i-1}^{*})$
or $\bm{\pi}^{*}\bm{a}_{i}^{l}=0$ for all $i=1,\cdots,Q_{1}$, and
the optimal threshold $i^{*}$ is zero. When $k_{1}\eta_{1}-k_{2}\eta_{2}<p_{max}<\tilde{p}_{0}$,
$g_{i-k_{1}}=f_{i}=1$ does not hold for all $i>0$ and hence there
must exist $i^{*}>0$.

\subsection{Proof of Theorem \ref{thm:opt_structure}\label{sub:Proof-of-Theorem6}}

\noindent Subject to the constraint $\sum_{i=0}^{Q_{1}}\pi_{i}=1$,
The average queueing delay $\bar{D}=\frac{1}{k_{1}\eta_{1}}\sum_{i=1}^{Q_{1}}i\pi_{i}$
can be minimized, if each $\pi_{i}$ chooses its lower bound $\Theta_{l}(i,\tilde{\bm{\pi}}_{i-1})$
for all $i\geq1$, corresponding to the case when $p_{max}\geq\tilde{p}_{0}$
shown in the proof of Theorem \ref{thm:opt_threshold}. Here, we focus
on studying the optimal solution in the case when $k_{1}\eta_{1}-k_{2}\eta_{2}<p_{max}<\tilde{p}_{0}$.
In this case, $\bar{P}=\sum_{i=0}^{Q_{1}}\xi_{i}\pi_{(i,0)}^{*}-\sum_{i=0}^{Q_{1}}\zeta_{i}\cdot\pi_{(i,1)}^{*}=p_{max}$
is straightforward. This lies in the fact that the data queue length
becomes smaller when more reliable energy can be exploited to transmit
backlogged data packets. 

In the sequel, we will show that when $k_{1}\eta_{1}-k_{2}\eta_{2}<p_{max}<\tilde{p}_{0}$,
the average queueing delay can also be minimized, if the optimal solution
$\bm{\pi}^{*}$ satisfies (\ref{eq:length_condition_for_optimality})
($\pi_{0}$, $\pi_{1}$, $\cdots$, $\pi_{i^{*}-1}$ are assigned
to their upper bounds and $\pi_{i^{*}+1},\cdots,\pi_{Q_{1}}$ are
assigned to their lower bounds) for Cases I, II, and III, respectively.

\subsubsection{Case I}

\noindent From Lemma \ref{lem:recurrent_transient_states}, we only
need to consider the steady-state probabilities $\pi_{(i,0)}$ and
$\pi_{(0,j)}$ for all $i\in\mathcal{Q}_{1}$ and $j\in\mathcal{Q}_{2}$.
The constraint $\sum_{j=0}^{Q_{2}}\pi_{(i,j)}\geq\Theta_{l}(i,\tilde{\bm{\pi}}_{i-1})=0$
naturally holds since $\pi_{(i,j)}\geq0$, and hence we do not consider
the corresponding constraints $\sum_{j=0}^{Q_{2}}\pi_{(i,j)}\geq0$
($i>0$). 

From the Markov chain in Fig.$\,$\ref{fig:markov_chain_onefig}(c),
the optimal solution $\pi_{(0,j)}^{*}$ satisfies $\mu_{2}\pi_{(0,j)}^{*}=\mu_{0}\pi_{(0,j-1)}^{*}$
for $j=1,\cdots,Q_{2}-1$ and $\eta_{1}\pi_{(0,Q_{2})}^{*}=\mu_{0}\pi_{(0,Q_{2}-1)}^{*}$
for $j=Q_{2}$. Hence, the probability that the data queue length
is zero is obtained as $\pi_{0}^{*}=\pi_{(0,0)}^{*}+\pi_{(0,0)}^{*}\sum\limits _{j=1}^{Q_{2}-1}\phi^{-j}+\frac{\pi_{(0,0)}^{*}}{\phi^{Q_{2}-1}\phi_{1}}=\alpha\pi_{(0,0)}^{*}$,
where $\alpha=\sum_{i=0}^{Q_{2}-1}\phi^{-i}+\phi^{-(Q_{2}-1)}\phi_{1}^{-1}$.
Thus, we have $\sum_{i=1}^{Q_{1}}\pi_{(i,0)}^{*}=1-\pi_{0}^{*}=1-\alpha\pi_{(0,0)}^{*}$.
And the normalized average power consumption from the RES is $\bar{P}=\mu_{2}\pi_{(0,0)}^{*}+(\mu_{2}-\mu_{0})\sum_{i=1}^{Q_{1}}\pi_{(i,0)}^{*}=\mu_{2}\pi_{(0,0)}^{*}+(\mu_{2}-\mu_{0})(1-\alpha\pi_{(0,0)}^{*})$,
from which we obtain $\pi_{(0,0)}^{*}=\frac{p_{max}-(\mu_{2}-\mu_{0})}{\mu_{2}-\alpha(\mu_{2}-\mu_{0})}$.
Hence, $\pi_{(0,0)}^{*}$ depends only on $p_{max}$. As shown in
the proof of Theorem \ref{thm:opt_threshold}, when $p_{max}\geq\tilde{p}_{0}$,
we have $\pi_{0}^{*}=\alpha\pi_{(0,0)}^{*}=1$ and thus $\pi_{(0,0)}^{*}=\frac{1}{\alpha}$.
When $p_{max}<\tilde{p}_{0}$, there must exist $\pi_{(0,0)}^{*}<\frac{1}{\alpha}$
and $\sum_{i=1}^{Q_{1}}\pi_{(i,0)}^{*}>0$. 

By contradiction, we will show that the optimal solution $\pi_{(i,0)}^{*}$
satisfies (\ref{eq:length_condition_for_optimality}): $\pi_{(i,0)}^{*}=\phi\pi_{(i-1,0)}^{*}$
for $i<i^{*}$, $0<\pi_{(i,0)}^{*}<\phi\pi_{(i-1,0)}^{*}$ for $i=i^{*}$,
and $\pi_{(i,0)}^{*}=0$ for $i>i^{*}$, since it leads to the minimum
average queueing delay. Suppose that there exists another set of steady-state
probabilities $\pi_{(i,0)}$: $\pi_{(i,0)}=\pi_{(i,0)}^{*}=\pi_{(0,0)}^{*}\phi^{i}$
for $0\leq i<m$, and $0<\pi_{(m,0)}<\phi\pi_{(m-1,0)}$ for $m\leq i\leq i_{1}$.
Subject to $\pi_{0}+\sum_{i=1}^{i_{1}}\pi_{(i,0)}=\pi_{0}^{*}+\sum_{i=1}^{i^{*}}\pi_{(i,0)}^{*}=1$,
there must exist $\pi_{(i,0)}<\pi_{(i,0)}^{*}$ for $m\leq i<i^{*}$
and $\sum_{i=i^{*}}^{i_{1}}\pi_{(i,0)}-\pi_{(i^{*},0)}^{*}=\sum_{i=m}^{i^{*}-1}(\pi_{(i,0)}^{*}-\pi_{(i,0)})$
with $i_{1}\geq i^{*}$. Thus, we have $i^{*}(\sum_{i=i^{*}}^{i_{1}}\pi_{(i,0)}-\pi_{(i^{*},0)}^{*})=i^{*}\sum_{i=m}^{i^{*}-1}(\pi_{(i,0)}^{*}-\pi_{(i,0)})$.
Hence, the corresponding average queueing delay $\bar{D}=\frac{1}{\eta_{1}}\sum_{i=1}^{i_{1}}i\cdot\pi_{(i,0)}$
satisfies
\[
\begin{array}{rl}
 & \bar{D}=\frac{1}{\eta_{1}}\sum_{i=1}^{i^{*}}i\cdot\pi_{(i,0)}^{*}-\frac{1}{\eta_{1}}\sum_{i=1}^{i^{*}-1}i\cdot(\pi_{(i,0)}^{*}-\pi_{(i,0)})\\
 & +\frac{1}{\eta_{1}}(\sum_{i=i^{*}}^{i_{1}-1}i\pi_{(i,0)}-i^{*}\pi_{(i^{*},0)}^{*})>\bar{D}^{*}.
\end{array}
\]
As a result, we can obtain the minimum average queueing delay when
the optimal solution $\bm{\pi}^{*}$ satisfies (\ref{eq:length_condition_for_optimality}).

\subsubsection{Case II}

Similar to Case I, we have $\Theta_{l}(i,\tilde{\bm{\pi}}_{i-1})=0$,
as listed in Table \ref{Table_upper_lower_bounds}. From the corresponding
Markov chain shown in Fig.$\,$\ref{fig:markov_chain_onefig}(d),
we have $\pi_{i}=\Theta_{l}(i,\tilde{\bm{\pi}}_{i-1})=0$ and $\pi_{(i,j)}=0$
for all $i>m$, if $\pi_{m}=\Theta_{l}(m,\tilde{\bm{\pi}}_{m-1})$
holds. We first show that $\pi_{i}=\Theta_{l}(i,\tilde{\bm{\pi}}_{i-1})$
does not hold for $i\leq i^{*}$. If the solution $\bm{\pi}$ satisfies
$0<\pi_{i}\leq\Theta_{u}(i,\tilde{\bm{\pi}}_{i})$ for $1<i<i_{1}$
and $\pi_{i_{1}}=\Theta_{l}(i_{1},\tilde{\bm{\pi}}_{i_{1}-1})$ for
some $i_{1}\leq i^{*}$, the corresponding power consumption from
the RES $\bar{P}$ will be larger than $p_{max}$, since it satisfies
$\bar{P}\geq\tilde{p}_{i_{1}-1}\geq\tilde{p}_{i^{*}-1}>p_{max}\geq\tilde{p}_{i^{*}}$
($\tilde{p}_{m}$ decreases with the threshold $m$). This violates
the constraint (\ref{eq:LP_problem}.a). Hence, the solution $\bm{\pi}^{*}$
should satisfy $0<\pi_{i}^{*}\leq\Theta_{u}(i,\tilde{\bm{\pi}}_{i}^{*})$
for $0<i\leq i_{1}$, and $\pi_{i}^{*}=\Theta_{l}(i,\tilde{\bm{\pi}}_{i-1}^{*})=0$
for $i\geq i_{1}>i^{*}$, respectively. Then, we show that among all
the candidate solutions, the solution $\bm{\pi}^{*}$ satisfying (\ref{eq:length_condition_for_optimality})
leads to the minimum queueing delay. According to the condition, we
have $0<\pi_{i}\leq\tau\bar{\eta}_{2}\pi_{(i-1,0)}+\pi_{(i,0)}\bar{\eta}_{2}$
for $0<i\leq i_{1}$. In this case, the solution $\bm{\pi}^{*}$ satisfies
$\pi_{i}^{*}=\sum_{j=0}^{Q_{2}}\pi_{(i,j)}^{*}=\tau\bar{\eta}_{2}\pi_{(i-1,0)}^{*}+\bar{\eta}_{2}\pi_{(i,0)}^{*}$
for $0<i<i^{*}$ and the minimum average queueing delay is $\bar{D}^{*}=\sum_{i=1}^{i^{*}}i\pi_{i}^{*}$.
Suppose that there is a solution $\bm{\pi}$ that satisfies $\pi_{i}=\tau\bar{\eta}_{2}\pi_{(i-1,0)}+\bar{\eta}_{2}\pi_{(i,0)}$
for $0<i<m$, $0<\pi_{m}<\tau\bar{\eta}_{2}\pi_{(m-1,0)}+\bar{\eta}_{2}\pi_{(m,0)}$
for some $m\leq i^{*}-1$, and $0<\pi_{i}\leq\tau\bar{\eta}_{2}\pi_{(i-1,0)}+\bar{\eta}_{2}\pi_{(i,0)}$
for $m<i\leq i_{1}$. Subject to $\sum_{i=0}^{i*}\pi_{i}^{*}=\sum_{i=0}^{i_{1}}\pi_{i}=1$,
the average queueing delay $\bar{D}=\frac{1}{\eta_{1}}\sum_{i=1}^{i_{1}}i\cdot\pi_{i}$
satisfies
\[
\begin{array}{l}
\bar{D}=\frac{1}{\eta_{1}}\sum_{i=1}^{i^{*}}i\cdot\pi_{i}^{*}+\frac{1}{\eta_{1}}(\sum_{i=1}^{i^{*}-1}i\cdot(\pi_{i}-\pi_{i}^{*})\\
+(\sum_{i=i^{*}}^{i_{1}}i\pi_{i}-i^{*}\pi_{i^{*}}^{*}))>\bar{D}^{*}+\frac{1}{\eta_{1}}\sum_{i=0}^{i^{*}-1}(i^{*}-i)(\pi_{i}^{*}-\pi_{i})\\
=\bar{D}^{*}+\frac{1}{\eta_{1}}\sum_{i=0}^{i^{*}-1}(i^{*}-i)\bar{\eta}_{2}(\pi_{(i,0)}^{*}-\pi_{(i,0)})\\
\quad+\frac{1}{\eta_{1}}\sum_{i=0}^{i^{*}-2}(i^{*}-i-1)\tau\bar{\eta}_{2}(\pi_{(i,0)}^{*}-\pi_{(i,0)})\geq\bar{D}^{*}
\end{array}
\]
where the first inequality holds since $\sum_{i=i^{*}}^{i_{1}}i\pi_{i}-i^{*}\pi_{i^{*}}^{*}=\sum_{i=0}^{i^{*}-1}i^{*}(\pi^{*}-\pi_{i})+\sum_{i=i^{*}+1}^{i_{1}}(i-i^{*})\pi_{i}>\sum_{i=0}^{i^{*}-1}i^{*}(\pi^{*}-\pi_{i})$,
the last two inequalities hold since we obtain $\pi_{0}>\bar{\eta}_{2}\pi_{(0,0)}$
and $\sum_{i=0}^{i^{*}-1}\pi_{(i,0)}^{*}\geq\sum_{i=0}^{i^{*}-1}\pi_{(i,0)}$
based on the property of the Markov chain. Hence, the optimal solution
$\bm{\pi}^{*}$ should satisfy (\ref{eq:length_condition_for_optimality})
to minimize the average queueing delay.

\subsubsection{Case III}

In this case, we have $\Theta_{l}(i,\tilde{\bm{\pi}}_{i-1})=\tau\sum_{m=[i-k_{1}+1]^{+}}^{i-1}\pi_{m}>0$.
Note that $Q_{1}$ should be sufficiently large to avoid buffer overflow.
To decrease the average queueing delay $\bar{D}=\frac{1}{k_{1}\eta_{1}}\sum_{i=1}^{Q_{1}}i\pi_{i}$,
$\pi_{i}$ with large index should be assigned its lower bound $\tau\sum_{m=[i-k_{1}+1]^{+}}^{i-1}\pi_{m}$.
In this sense, there exists an integer $i_{1}$ that satisfies $\pi_{i}=\tau\sum_{m=[i-k_{1}+1]^{+}}^{i-1}\pi_{m}$
for all $i\geq i_{1}$. For the same reason as stated in Case II,
$i_{1}>i^{*}$ should be satisfied to meet the power consumption (from
the RES) constraint $\bar{P}\leq p_{max}$. 

In the same way, we will compare the delay performances between the
optimal solution $\bm{\pi}^{*}$ satisfying (\ref{eq:length_condition_for_optimality})
and a candidate solution $\bm{\pi}$. Suppose that $\bm{\pi}$ satisfies
$\pi_{i}=\Theta_{u}(i,\tilde{\bm{\pi}}_{i})$ for $0<i<m$, $\Theta_{l}(i,\tilde{\bm{\pi}}_{i-1})<\pi_{m}<\Theta_{u}(i,\tilde{\bm{\pi}}_{i})$
for some $m\leq i^{*}-1$, and $0<\pi_{i}\leq\Theta_{u}(i,\tilde{\bm{\pi}}_{i})$
for $m<i\leq i_{1}$. We notice that $\{\pi_{i}^{*}=\tau\sum_{m=[i-k_{1}+1]^{+}}^{i-1}\pi_{m}^{*}\}$
($i>i^{*}$) is a decreasing sequence (otherwise the data queue will
be unstable). So does the sequence $\{\pi_{i}\}$ ($i>i_{1}$). And
$\pi_{i}^{*}=\Theta_{u}(i,\tilde{\bm{\pi}}_{i}^{*})$ increases with
the growth of $i$ for $0\leq i\leq i^{*}$. Then, subject to the
constraint $\sum_{i=0}^{Q_{1}}\pi_{i}^{*}=\sum_{i=0}^{Q_{1}}\pi_{i}=1$,
we have $\pi_{i}^{*}\geq\pi_{i}$ for $0\leq i\leq i^{*}$ and $\pi_{i}^{*}\leq\pi_{i}$
for $i^{*}<i\leq Q_{1}$, and $\sum_{i=0}^{i^{*}}(\pi_{i}^{*}-\pi_{i})=\sum_{i=i^{*}+1}^{Q_{1}}(\pi_{i}-\pi_{i}^{*})$.
As a result, the average queueing delay $\bar{D}$ satisfies 
\[
\begin{array}{l}
\bar{D}=\frac{1}{k_{1}\eta_{1}}\sum\limits _{i=1}^{Q_{1}}i\pi_{i}^{*}+\frac{1}{k_{1}\eta_{1}}(\sum\limits _{i=1}^{i^{*}}i(\pi_{i}-\pi_{i}^{*})+\sum\limits _{i=i^{*}+1}^{Q_{1}}i(\pi_{i}-\pi_{i}^{*}))\\
>\bar{D}^{*}+\frac{1}{k_{1}\eta_{1}}(-\sum\limits _{i=1}^{i^{*}}i^{*}(\pi_{i}-\pi_{i}^{*})+\sum\limits _{i=i^{*}+1}^{Q_{1}}i(\pi_{i}-\pi_{i}^{*}))>\bar{D}^{*}.
\end{array}
\]
In this way, we show that the optimal solution $\bm{\pi}^{*}$ should
satisfy (\ref{eq:length_condition_for_optimality}) in Case III. 

Note that the optimal solution $\bm{\pi}^{*}$ corresponds to the
threshold based scheduling policy. Naturally, a larger threshold $m$
leads to a larger queueing delay. Meanwhile, a lower power $\tilde{p}_{m}$
is consumed from the RES. The optimal threshold can be obtained by
comparing the maximum allowable power consumption $p_{max}$ with
the power thresholds $\{\tilde{p}_{m}\}$ as: $i^{*}=\arg\min_{\tilde{p}_{m}\leq p_{max}}m$.

\subsection{Proof of Corollary \ref{cor:opt_solution_SA}\label{sub:Proof-of-Corollary8}}

\noindent From Theorem \ref{thm:opt_structure} and Lemma \ref{lem:relation_pi},
when $p_{max}\geq\tilde{p}_{0}$, $\pi_{i}^{*}=\pi_{(i,0)}^{*}=0$
for all $i>0$. Then, by substituting (\ref{eq:pi_0j_star}) into
the equation $\pi_{(0,0)}^{*}+\sum_{j=1}^{Q_{2}}\pi_{(0,j)}^{*}=1$,
we obtain $\pi_{(0,0)}^{*}=\frac{1}{\alpha}$. Accordingly, the power
threshold $\tilde{p}_{0}=\mu_{2}\pi_{(0,0)}^{*}=\mu_{2}\alpha^{-1}$
because of $g_{0}=1$. Then, we discuss the optimal solution for the
case when $\eta_{1}-\eta_{2}<p_{max}<\tilde{p}_{0}$. From Theorem
\ref{thm:opt_structure} and its proof, we know that there exists
an optimal threshold $i^{*}>0$ so that $\pi_{(i,0)}^{*}=\phi\pi_{(i-1,0)}^{*}=\pi_{(0,0)}^{*}\phi^{i}$
for $i<i^{*}$ and $\pi_{(i,0)}^{*}=0$ for $i>i^{*}$, where $\pi_{(0,0)}^{*}=\frac{p_{max}-(\mu_{2}-\mu_{0})}{\mu_{2}-\alpha(\mu_{2}-\mu_{0})}$.
Since $\pi_{0}^{*}=\alpha\pi_{(0,0)}^{*}$ and thus $\sum_{i=1}^{i^{*}}\pi_{(i,0)}^{*}=1-\alpha\pi_{(0,0)}^{*}$,
we obtain $\pi_{(i^{*},0)}^{*}=1-\alpha\pi_{(0,0)}^{*}-\sum_{i=1}^{i^{*}-1}\pi_{(i,0)}^{*}=1-\pi_{(0,0)}^{*}(\alpha+\sum_{i=1}^{i^{*}-1}\phi^{i})$.
From the property of the optimal solution $\pi_{(i,0)}^{*}$, the
optimal threshold $i^{*}$ can be evaluated as $i^{*}=\Omega_{\phi}(\pi_{(0,0)}^{*},1-\alpha\pi_{(0,0)}^{*})$,
which is the integer that satisfies $\pi_{(0,0)}^{*}\sum_{m=1}^{i^{*}-1}\phi^{m}\leq1-\alpha\pi_{(0,0)}^{*}<\pi_{(0,0)}^{*}\sum_{m=1}^{i^{*}}\phi^{m}$.

\subsection{Proof of Corollary \ref{cor:opt_solution}\label{sub:Proof-of-Corollary9}}

\noindent From Theorem \ref{thm:opt_structure} and its proof, the
optimal threshold is $i^{*}=0$ when $p_{max}\ge\tilde{p}_{0}$ and
hence the optimal solution $\bm{\pi}^{*}$ can be obtained by solving
$(1+Q_{1})(1+Q_{2})$ independent linear equations: $\bm{\pi}\bm{a}_{i}^{u}=0$
$(\forall i>0)$, $\bm{\pi}\tilde{\bm{P}}_{s}=\bm{0}$, and $\bm{\pi}\bm{e}=1$.
In this case, we get $\bm{\pi}^{*}=\bm{\pi}_{0}^{'}$ according to
(\ref{eq:opt_pi_m}). When $\tilde{p}_{i^{*}}\leq p_{max}<\tilde{p}_{i^{*}-1}$,
the optimal solution $\bm{\pi}^{*}$ satisfies $\bm{\pi}^{*}\bm{a}_{0}=p_{max}$,
$\bm{\pi}^{*}\bm{a}_{i}^{u}=0\,(i=1,\cdots,i^{*}-1)$, $\bm{\pi}^{*}\bm{a}_{i}^{l}=0\,(i=i^{*}+1,\cdots,Q_{1})$,
$\bm{\pi}\tilde{\bm{P}}_{s}=\bm{0}$, and $\bm{\pi}\bm{e}=1$. Hence,
we can obtain $\bm{\pi}^{*}=\bm{b}\bm{A}^{-1}$ by solving $(1+Q_{1})(1+Q_{2})$
linear equations.

\subsection{Proof of Corollary \ref{cor:opt_gi_fi}\label{sub:Proof-of-Corollary10}}

We will discuss the optimal transmission parameters in Cases I, II
and III, respectively.

\subsubsection{Case I}

When $p_{max}\geq\tilde{p}_{0}$, from the local equilibrium equation
$\pi_{(0,0)}^{*}\lambda_{1,0}^{*}=\pi_{(1,0)}^{*}\tilde{\mu}_{1,1}^{*}=0$,
we must have $\lambda_{1,0}^{*}=\mu_{2}(1-g_{0}^{*})=0$ and $g_{0}^{*}=1$,
since $\pi_{(0,0)}^{*}=\alpha^{-1}>0$ and $\pi_{(i,0)}^{*}=0$ for
all $i>0$. Also, since $\pi_{(i,0)}^{*}\lambda_{1,i}^{*}=\pi_{(i+1,0)}^{*}\tilde{\mu}_{1,i+1}^{*}=0$
always holds for $i>0$, we can set $g_{i}^{*}=1$ for all $i\geq0$
and $f_{i}^{*}=0$ for $i>0$. When $\eta_{1}-\eta_{2}<p_{max}<\tilde{p}_{0}$,
from (\ref{eq:opt_prob_pi_1i}), $\pi_{(i,0)}^{*}=\pi_{(0,0)}^{*}\phi^{i}$
for all $i<i^{*}$. Thus, $\frac{\pi_{(i+1,0)}^{*}}{\pi_{(i,0)}^{*}}=\frac{\lambda_{1,i}^{*}}{\tilde{\mu}_{1,i+1}^{*}}=\phi$.
On the other hand, $0\leq\lambda_{1,i}^{*}\leq\mu_{2}$ and $\mu_{0}\leq\tilde{\mu}_{1,i}^{*}\leq\bar{\eta}_{1}$.
Hence, we have $\lambda_{1,i}^{*}=\mu_{2}$ and $\tilde{\mu}_{1,i+1}^{*}=\mu_{0}$
for $i<i^{*}-1$. Substituting $\lambda_{1,i}^{*}$ and $\tilde{\mu}_{1,i+1}^{*}$
into (\ref{eq:par_mu_1i_case1}) gives $g_{i}^{*}=0$ for $0\leq i\leq i^{*}-2$
and $f_{i}^{*}=0$ for $1\leq i\leq i^{*}-1$. 

From $\pi_{(i-1,0)}^{*}\lambda_{1,i-1}=\pi_{(i,0)}^{*}\tilde{\mu}_{1,i}$,
we can get $\lambda_{1,i^{*}}^{*}=0$ and $g_{i^{*}}^{*}=1$, since
$\pi_{(i,0)}^{*}=0$ for all $i>i^{*}$. Without loss of generality,
we can set $g_{i}=1$ and $f_{i}=0$ for $i>i^{*}$ to maintain consistency.
From $\pi_{(i^{*}-1,0)}^{*}\lambda_{1,i^{*}-1}=\pi_{(i^{*},0)}^{*}\tilde{\mu}_{1,i^{*}}$,
we have $\mu_{2}(1-g_{i^{*}-1}^{*})=\frac{\pi_{(i^{*},0)}^{*}}{\pi_{(i^{*}-1,0)}^{*}}(\mu_{0}+\mu_{1}f_{i^{*}})$,
which is satisfied when $f_{i^{*}}=0$ and $g_{i^{*}-1}^{*}=1-\frac{\mu_{0}}{\mu_{2}}\frac{\pi_{(i^{*},0)}^{*}}{\pi_{(0,0)}^{*}\phi^{i^{*}-1}}=1-\frac{\pi_{(i^{*},0)}^{*}}{\pi_{(0,0)}^{*}\phi^{i^{*}}}$.
In this way, we get $f_{i}^{*}=0$ for all $i>0$ and $g_{i}^{*}$
as listed in Table \ref{Table_optimal_parameters}.

\subsubsection{Case II}

Similar to Case I, by exploiting the local equilibrium equation $\pi_{(i-1,0)}^{*}\lambda_{1,i-1}^{*}=\pi_{(i,0)}^{*}\mu_{1,i}^{*}+\bar{\eta}_{1}\sum_{j=1}^{Q_{2}}\pi_{(i,j)}^{*}$
and $\pi_{(i,j)}^{*}=0$ for all $i>i^{*}$, we can obtain $g_{i}^{*}=1$
($i\geq0$) and $f_{i}^{*}=0$ ($i>0$) when $p_{max}\geq\tilde{p}_{0}$,
and $f_{i}^{*}=0$ and $g_{i}^{*}$ listed in Table \ref{Table_optimal_parameters}
when $\eta_{1}-k_{2}\eta_{2}<p_{max}<\tilde{p}_{0}$, respectively.

\subsubsection{Case III}

Similar to the above two cases, when $p_{max}\geq\tilde{p}_{0}$,
we have $g_{i}^{*}=f_{i+1}^{*}=1$ for all $i\geq0$, since $i^{*}=0$.
When $k_{1}\eta_{1}-\eta_{2}<p_{max}<\tilde{p}_{0}$, the local equilibrium
equation at the state $(i^{*},0)$ is given by $\pi_{(i^{*},0)}^{*}(\mu_{0}+\mu_{1}f_{i^{*}}^{*})-\pi_{(i^{*}-k_{1},0)}^{*}\mu_{2}(1-g_{i^{*}-k_{1}}^{*})=\eta_{1}\sum_{m=i^{*}-k_{1}+1}^{i^{*}-1}\sum_{j=0}^{Q_{2}}\pi_{(m,j)}^{*}-\bar{\eta}_{1}\sum_{j=1}^{Q_{2}}\pi_{(i^{*},j)}^{*}$
when $i^{*}\geq k_{1}$. When $0\leq i^{*}<k_{1}$, the local equilibrium
equation at the state $(i^{*},0)$ can be expressed as $\pi_{(i^{*},0)}^{*}(\mu_{0}+\mu_{1}f_{i^{*}}^{*})=\eta_{1}\sum_{m=0}^{i^{*}-1}\sum_{j=0}^{Q_{2}}\pi_{(m,j)}^{*}-\bar{\eta}_{1}\sum_{j=1}^{Q_{2}}\pi_{(i^{*},j)}^{*}$.
From the local equilibrium equations, we can compute $g_{i^{*}-k_{1}}^{*}$
and $f_{i^{*}}^{*}$ for $i^{*}\geq k_{1}$ and $i^{*}<k_{1}$, respectively,
as listed in Table \ref{Table_optimal_parameters}. 

\bibliographystyle{IEEEtran}
\bibliography{IEEEabrv,IEEEexample,PaperLib}

\end{document}